\documentclass[12pt,subeqn,a4paper]{article}

\usepackage{amssymb,amsmath,amsfonts,amsthm, amscd, mathrsfs,helvet,multirow}
\usepackage{bm}
\usepackage{graphicx,verbatim}
\usepackage{psfrag}
\usepackage[all]{xy}

\usepackage{appendix}
\numberwithin{equation}{section}
\numberwithin{figure}{section}
\newcommand{\N}{\mathcal{N}}
\newcommand{\M}{\mathcal{M}}
\newcommand{\F}{\mathcal{F}}
\DeclareMathOperator{\im}{Im}
\DeclareMathOperator{\re}{Re}
\newcommand{\inv}{\left(\im \tau^{-1}\right)}

\def\r2{\sqrt{2}}

\begin{document}
\newcommand{\nd}[1]{/\hspace{-0.5em} #1}
\begin{titlepage}
\begin{flushright}
DAMTP-2014-57\\
\end{flushright}
\begin{centering}
\vspace{.2in}
 {\large {\bf Superconformal Quantum Mechanics and the Discrete 
     Light-Cone Quantisation of {\cal N}=4 SUSY Yang-Mills }}

\vspace{.3in}

Nick Dorey and Andrew Singleton\\
\vspace{.1 in}
DAMTP, Centre for Mathematical Sciences \\ 
University of Cambridge, Wilberforce Road \\ 
Cambridge CB3 0WA, UK \\
\vspace{.2in}
%
%
\vspace{.4in}
{\bf Abstract} \\

\end{centering}
We study the quantum mechanical $\sigma$-model arising in the 
discrete light-cone quantisation of ${\cal N}=4$ supersymmetric
Yang-Mills theory. The target space is a certain torus fibration over a
scale-invariant special K\"{a}hler manifold. We show that the expected 
$SU(1,1|4)$ light-cone superconformal invariance of the ${\cal N}=4$
theory emerges in a limit where the volume of the fibre goes to zero
and give an explicit construction of the generators. The construction 
given here yields a large new family of superconformal quantum
mechanical models with $SU(1,1|4)$ invariance.

\end{titlepage}
\paragraph{}
\section{Introduction}

Supersymmetric gauge theories in four dimensions exhibit a wide range
of interesting phenomena including a remarkable web of dualities
relating seemingly different models. Many of these properties can be
understood in terms of a mysterious conformal theory with $(2,0)$
supersymmety in six
dimensions which gives rise to various 4d gauge theories after
compactification \cite{Witten:1995}. 
The simplest case is the compactification of this 
$(2,0)$ theory on a two-torus which gives ${\cal N}=4$ supersymmetric
gauge theory in four dimensions at low energy. Modular transformations
of the torus then correspond to electric-magnetic duality
transformations in the low-energy theory. 
\paragraph{}
Although we know very little
about the $(2,0)$ theory, a concrete proposal \cite{Aharonyetal:1997} 
exists to define the theory 
compactified on a null circle in terms of quantum mechanics on
the moduli space of Yang-Mills instantons. This is an example of the more
general phenomenon of discrete light-cone quantisation (DLCQ), where
restriction to a sector of fixed null momentum yields a finite
dimensional quantum mechanical model. The relation of the
six-dimensional theory to ${\cal N}=4$ super-Yang-Mills in four
dimensions also gives rise to a related proposal for the DLCQ
description of the latter theory \cite{Ganor:Sethi:1997,Kapustin:Sethi:1998}. The main goal of this
paper is to formulate the quantum
mechanical model describing the DLCQ of the ${\cal N}=4$ theory 
explicitly and construct its symmetry algebra. 
\paragraph{}           
In discrete light-cone quantisation, the quantum mechanical model
describing a sector of fixed null momentum inherits a subgroup of the
spacetime symmetry of the full theory. In the case of a superconformal
field theory, the reduced model is invariant under the subgroup of the 
superconformal symmetry which commutes with the null momentum. As we
review in Section \ref{section:N=4DLCQ} below, this corresponds to an $SU(1,1|4)$
subgroup of the $PSU(2,2|4)$ superconformal invariance of the ${\cal
  N}=4$ theory. Thus we seek an $SU(1,1|4)$-invariant superconformal
quantum mechanics. 
The model proposed in \cite{Ganor:Sethi:1997, Kapustin:Sethi:1998} takes the form of a quantum
mechanical $\sigma$-model. As we review in Section \ref{section:6dDLCQ} below, the target
space corresponds to the moduli space of instantons in an auxiliary
Yang-Mills theory living on $\mathbb{R}^{2}\times T^{2}$. More
precisely we should consider a limit where the area of the the torus
goes to zero. We will show that the expected $SU(1,1|4)$ invariance
indeed emerges in this limit.  
\paragraph{}
A beautiful feature of non-linear $\sigma$-models is that the
conditions for unbroken supersymmetry have a geometric character. 
A famous example is that ${\cal
    N}=(4,4)$ supersymmetry in one or two dimensions requires a hyper
  K\"{a}hler target \cite{AlvarezGaume:Freedman:1981}.    
Similarly, superconformal 
invariance of a quantum mechanical $\sigma$-model also constrains the
geometry of the target
space \cite{Michelson:Strominger:1999}. Although 
several families of superconformal $\sigma$-models corresponding to
different target space geometries are known, the conditions for  
$SU(1,1|4)$ invariance have not been discussed before in the
literature. Our main result is that scale-invariant target spaces with 
{\em special K\"{a}hler geometry} naturally solve these constraints. 
As we discuss below, the dimensional reduction of ${\cal N}=2$ 
superconformal models in four dimensions provides a large class of
interesting examples. In particular, 
the relevant target space for the DLCQ description of
the ${\cal N}=4$ theory takes the form of a torus fibration over a 
special K\"{a}hler manifold.  
In the limit relevant for ${\cal N}=4$ super-Yang-Mills, the volume of
the fibre goes to zero and the model is one of this class.  
\paragraph{}  
The paper is organised as follows. After a brief review of discrete
light-cone quantisation and its application to the ${\cal N}=4$
theory, we proceed to construct a quantum mechanical $\sigma$-model
whose target space is a certain torus bundle over a generic
scale-invariant special K\"{a}hler
base. Although our main interest is the limit in which the dynamics in
the fibre directions decouples giving superconformal quantum
mechanics on the base, the more general model is interesting for two 
reasons. First, it arises in the context of DLCQ where it
corresponds to the compactification of the $(2,0)$ theory on a torus
of finite area. Second, it provides a natural setting for the
resolution of the singularities of the base manifold which is likely
needed to make sense of these models. The bulk of the paper is devoted
to studying the symmetry algebra of the full $\sigma$-model and its
enhancement to $SU(1,1|4)$ in the relevant limit. A brief discussion of
singularities and their resolution is given in the final section. In
this paper, we focus mainly on the general class of models described
above. Discussion of the particular case relevant to the DLCQ of the
${\cal N}=4$ theory will be given in a separate paper \cite{ND:AS:future}.

\section{${\cal N}$=4 SUSY Yang-Mills on the Light-Cone}
\label{section:N=4DLCQ}

We will consider ${\cal N}=4$ supersymmetric Yang-Mills theory with
gauge group $SU(N)$ and complexified coupling 
\begin{eqnarray}
\tau & = & \frac{4\pi i}{g^{2}}\,+\, \frac{\theta}{2\pi}.
\nonumber 
\end{eqnarray}
The theory defined on $\mathbb{R}^{3,1}$ has superconformal invariance
which is unbroken at the origin of the moduli space where the vacuum
expectation values of all scalar fields vanish. The full
superconformal group is isomorphic to $PSU(2,2|4)$. In addition to the
usual Poincar\'{e} generators $P_{m}$, $M_{mn}$, 
the bosonic part of the  
corresponding algebra includes the dilatation operator $D$ and special
conformal transformations $K_{m}$ as well as the generators
$R^{A\bar{B}}$ of the $SU(4)$ R-symmetry. The algebra is completed by
Poincar\'{e} supersymmetry generators $Q^{A}_{\alpha}$, 
$\bar{Q}^{\bar{A}}_{\dot{\alpha}}$ in the 
$({\bf 2},{\bf 1},{\bf 4})\oplus({\bf 1},{\bf 2},\bar{\bf 4})$ of
$SU(2)_{L}\times SU(2)_{R}\times SU(4)$ and superconformal generators 
$S^{A}_{\alpha}$, 
$\bar{S}^{\bar{A}}_{\dot{\alpha}}$ also transforming as 
$({\bf 2},{\bf 1},{\bf 4})\oplus({\bf 1},{\bf 2},\bar{\bf 4})$.         
\paragraph{}
In discrete light-cone quantisation (DLCQ) one considers the theory
compactified on a light-like circle. Starting from Minkowski space with
Cartesian coordinates $\{x_{0},x_{1},x_{2},x_{3}\}$, we define 
light-cone coordinates $x_{+}=x_{0}+x_{1}$ and $x_{-}=x_{0}-x_{1}$ and
impose the periodic identification $x_{-}\sim x_{-}+2\pi R_{-}$. Thus
we replace Minkowski space with the spacetime manifold 
\begin{eqnarray}
{\cal M}_{4} & = & \mathbb{R}^{2}\times S^{1}_{-}\times \mathbb{R}_{+}, 
\nonumber 
\end{eqnarray}
where $x_{\pm}$ are coordinates on  $\mathbb{R}_{+}$ and $S^{1}_{-}$
respectively.  
The momentum $p_{+}=p_{0}+p_{1}$ conjugate to $x_{-}$, which is manifestly
positive for on-shell states, is thus quantised as 
$p_{+}=K/R_{-}$ where $K$ is a positive integer. In light-cone
quantisation the coordinate $x_{+}$ plays the role of time and the
conjugate momentum $p_{-}=p_{0}-p_{1}$ is the corresponding
Hamiltonian. Working in a sector of fixed $K$ typically reduces the
field theory to a finite-dimensional quantum mechanics model.  
\paragraph{}
Although light-like compactification on a circle of
fixed radius breaks dilatation symmetry of the four-dimensional
theory, a linear combination of scaling with 
a Lorentz boost in the compact direction remain unbroken. The
corresponding generator  $T=D+M_{01}$ is known as the light-cone
dimension. More generally, 
compactification on ${\cal M}_{4}$ breaks the four-dimensional superconformal
group down to the ``collinear subgroup''\footnote{More precisely the
  collinear subgroup described eg in \cite{Belitskyetal:2004} contains an 
extra generator which preserves the light-cone but not the radius of
the light-like compactification.}. 
The four-dimensional conformal group is broken as, 
\begin{eqnarray}
SO(4,2) & \rightarrow& SO(2,1) \times SO(2) 
\nonumber 
\end{eqnarray}
where the unbroken $SO(2,1)$ factor is generated by the light-cone
dimension $T=D+M_{01}$, the light-cone Hamiltonian $H=p_{-}$ and the
special conformal transformation $K=K_{0}+K_{1}$;
\begin{eqnarray}
[T,K]\,=\,2iK \qquad{}  & [T,H]\,=\,-2iH  & \qquad{} [H,K]=-4iT.
\nonumber 
\end{eqnarray}
The unbroken $SO(2)$, with generator $J=M_{23}$, corresponds to
rotations in the transverse $\mathbb{R}^{2}$. In fact the
DLCQ theory has a larger spacetime symmetry group known as the
Schr\"{o}dinger group which also contains Galilean boosts, but
 these extra generators do not affect the following and we will
not discuss them.   
\paragraph{}
Light-like compactification also breaks half of the fermionic
symmetries of the four-dimensional theory. In particular each of the 
two-component spinor
supercharges  $Q^{A}_{\alpha}$, 
$\bar{Q}^{\bar{A}}_{\dot{\alpha}}$, $S^{A}_{\alpha}$ and  
$\bar{S}^{\bar{A}}_{\dot{\alpha}}$ has a projection onto the light-cone 
corresponding to an unbroken symmetry (see \cite{Belitskyetal:2004} for further
details). We denote the unbroken
generators as   $Q^{A}$, 
$\bar{Q}^{\bar{A}}$, $S^{A}$ and  
$\bar{S}^{\bar{A}}$. The $SU(4)$ R-symmetry also remains
unbroken and the full unbroken bosonic symmetry is therefore 
\begin{eqnarray}         
G_{B} \,=\, SO(2,1)\times SO(2) \times SU(4) & \simeq  &  SU(1,1)\times U(4).
\nonumber 
\end{eqnarray}
The fermionic generators $Q^{A}$, $S^{A}$ form a doublet of $SU(1,1)$
as do $\bar{Q}^{\bar{A}}$,  $\bar{S}^{\bar{A}}$. 
\paragraph{}
The full symmetry of DLCQ 
should correspond to a Lie superalgebra whose maximal bosonic subalgebra
is ${\rm Lie}(G_{B})$ and which also includes sixteen fermionic
generators in the $({\bf 2},{\bf 4})\oplus ({\bf 2},\bar{\bf 4})$ of 
$G_{B}$. Up to automorphisms, the unique possibility is $SU(1,1|4)$. 
The DLCQ of ${\cal N}=4$ SUSY Yang-Mills should therefore be
superconformal quantum mechanics with this symmetry. The main
goal of this paper is to identify this model and construct its
symmetry generators explicitly.   
\paragraph{}
The light-like compactification described above also allows one to
introduce Wilson lines for the $SU(N)$ gauge field 
\begin{eqnarray}
\left\langle \oint_{S^{1}_{-}}\, A\cdot dx \right\rangle & = & 
{\rm diag}\left\{\mu_{1},\mu_{2},\ldots,\mu_{N}\right\},
\qquad{} \qquad{} \sum_{i=1}^{N} \,\mu_{i}\,=\,0. \nonumber 
\end{eqnarray}
If $\mu_{i}\neq\mu_{j}$ for all $i$ and $j$ then the gauge group is
broken down to its Cartan subalgebra by the adjoint Higgs mechanism: 
\begin{eqnarray} 
SU(N) & \rightarrow & U(1)^{N-1}. 
\nonumber
\end{eqnarray} 
Further, performing a duality transformation on the resulting
three-dimensional abelian low-energy effecive theory, one can also
introduce corresponding magnetic Wilson lines denoted $\rho_{i}$ for
$i=1,2,\ldots,N$. The electric and magnetic Wilson lines 
are naturally combined to form $N$ complex parameters,
$Z_{i}=\rho_{i}+\tau\mu_{i}$. Taking into account
the standard $2\pi$-periodicity of the Wilson lines, the $\{Z_{i}\}$ correspond
to $N$ points on a torus of complex structure $\tau$. For gauge group
$SU(N)$, only the $N-1$ relative positions of these points are 
significant.      
\paragraph{}
Importantly neither the electric or magnetic Wilson
lines break the light-cone superconformal symmetry of DLCQ identified
above. Thus we seek a quantum mechanical model with $SU(1,1|4)$
superconformal symmetry and $N-1$ additional complex parameters. 

\section{ DLCQ of 
${\cal N}$=4 SUSY Yang-Mills from Six Dimensions}
\label{section:6dDLCQ}
\paragraph{}
The approach to the ${\cal N}=4$ theory which we take here starts by
realising the theory as a compactification of six-dimensional
conformal field theory. In particular ${\cal N}=4$ supersymmetric
Yang-Mills theory with gauge group $SU(N)$ arises as a low-energy
effective theory when the $(2,0)$ superconformal field theory of type
$A_{N-1}$ is compactified down to four dimensions on a two-dimensional 
torus \cite{Witten:1995}. 
The complex structure parameter of the torus coincides with the
complexified coupling $\tau=4\pi i/g^{2}+\theta/2\pi$ of the ${\cal
  N}=4$ theory. If the torus has area ${\cal A}$ the full theory also
contains an infinite tower of Kaluza-Klein (KK) modes corresponding to
states carrying momentum along the two compact dimensions. In the
limit ${\cal A}\rightarrow 0$ the KK modes decouple and the remaining
theory is precisely ${\cal N}=4$ super-Yang-Mills. 
\paragraph{}   
The $(2,0)$ theory in six non-compact dimensions 
has a well established DLCQ description 
\cite{Aharonyetal:1997,Aharony:Berkooz:Seiberg:1997}. 
Following our discussion of the four-dimensional theory above, we
compactify the $(2,0)$ theory on  
\begin{eqnarray}
{\cal M}_{6} & = & \mathbb{R}^{4}\times S^{1}_{-}\times \mathbb{R}_{+}, 
\nonumber 
\end{eqnarray}   
where $S^{1}_{-}$ is a light-like circle of radius $R_{-}$. The sector
of the theory with $K$ units of momentum in the compact direction is
described by supersymmetric quantum mechanics on the
moduli space of $K$ Yang-Mills instantons of gauge group $SU(N)$ on 
$\mathbb{R}^{4}$. The 
instanton moduli space is a hyper-K\"{a}hler manifold of real dimension $4KN$. 
A quantum mechanical $\sigma$-model with a hyper-K\"{a}hler target
admits an ${\cal N}=(4,4)$ supersymmetric extension. In fact the
instanton moduli space is also  
equipped with a triholomorphic homothety of degree two. Under these
conditions ${\cal N}=(4,4)$ supersymmetry is enlarged to give an 
$OSp(4|4)$ superconformal invariance \cite{AS:2014}. 
The latter coincides with the
subgroup of the $(2,0)$ superconformal algebra in six dimensions left
unbroken by compactification on ${\cal M}_{6}$.    
\paragraph{}
To obtain a DLCQ description of the ${\cal N}=4$ theory it is
necessary to compactify two of the transverse dimensions on a torus.  
Thus we consider the $(2,0)$ theory compactified on    
\begin{eqnarray}
\tilde{\cal M}_{6} & = & \mathbb{R}^{2}\times T^{2}_{\tau}\times S^{1}_{-}
\times \mathbb{R}_{+}. 
\nonumber 
\end{eqnarray} 
As above the complex structure parameter of the torus, denoted $\tau$,
is identified with the complexified coupling of the four-dimensional
gauge theory. The resulting description of the sector with $K$ units
of momentum along $S^{1}_{-}$ is again a quantum mechanical $\sigma$-model with
${\cal N}=(4,4)$ supersymmetry. The model, which was 
introduced in \cite{Ganor:Sethi:1997,Kapustin:Sethi:1998}, has as its target space the moduli space
of $K$ instantons in $SU(N)$ Yang-Mills theory living on   
$\mathbb{R}^{2}\times T^{2}_{\tau}$. In the following we will denote
this manifold as ${\cal M}_{K,N}$. 
To obtain the DLCQ of the ${\cal
  N}=4$ theory we should take the area ${\cal A}$ of the torus to zero
holding its shape fixed.      
\paragraph{}
The moduli space ${\cal M}_{K,N}$ is again a 
hyper-K\"{a}hler manifold of real dimension\footnote{More
  precisely, as we discuss below, the metric becomes singular along $4N-4$
  real directions.} $4KN$. Although, for general values
of the parameters ${\cal A}$ and $\tau$, the hyper-K\"{a}hler metric
is not known explicitly, the manifold ${\cal M}_{K,N}$ has (at least) two useful
descriptions. The first description arises via the ADHM Nahm transform 
which maps ${\cal M}_{K,N}$ to the moduli space of Hitchin's equations
on $\hat{T}^{2}_{\tau}$ in the presence of punctures at the points
$z=Z_{i}$, for $i=1,2,\ldots,N$, 
corresponding to the electric and magnetic Wilson lines discussed in
the previous section. In more
physical language, the moduli space can be thought of as the Higgs branch of an
auxiliary supersymmetric gauge theory on $\hat{T}^2_{\tau}$ with localised
impurities at the punctures \cite{Kapustin:Sethi:1998}. 
\paragraph{}
The second description of ${\cal
  M}_{K,N}$ arises via the three-dimensional mirror symmetry 
\cite{Intriligator:Seiberg:1996} which
maps the Higgs branch of the impurity theory to the Coulomb branch of
yet another auxiliary supersymmetric gauge theory \cite{Kapustin:1998}. 
The starting point for the Coulomb branch description is a
four-dimensional quiver gauge theory with ${\cal N}=2$
supersymmetry. The quiver diagram for the theory in question 
coincides with the Dynkin diagram for the affine Lie algebra
$\hat{A}_{N-1}$. The gauge group is 
\begin{eqnarray}
\hat{G} & = & U(K)_{1}\,\times \, U(K)_{2} \,\times 
\,\ldots \,\times \, U(K)_{N}.
\nonumber 
\end{eqnarray}
\paragraph{}
In addition to an ${\cal N}=2$ vector multiplet for each $SU(K)$
factor in $\hat{G}$, the theory contains hypermultiplets in the
bifundamental representation of adjacent factors. Thus we have a
hypermultiplet in the $(\bar{\bf k}, {\bf k})$ of  
$U(K)_{i}\times U(K)_{i+1}$ for $i=1,2,\ldots,N$ with the
identification $U(K)_{N+1}\simeq U(K)_{1}$. An important subtlety is
that the gauge coupling of the $U(1)$ center of each of 
the $N-1$ off-diagonal $U(K)$ factors in
$\hat{G}$ has a positive $\beta$ function and therefore exhibits a
Landau pole. The effect is to freeze out each of these $U(1)$ factors
to give a theory with gauge group  
\begin{eqnarray}
\hat{G}' & = & U(1)_{D}\times \prod_{j=1}^{N} \, SU(K)_{j},
\label{eqn:quivergroup} 
\end{eqnarray}     
where $U(1)_{D}$ corresponds to the center of the diagonal $U(K)$ in
the original gauge group $\hat{G}$. The $\beta$ functions for the
remaining gauge couplings vanish and the resulting theory is an 
${\cal N}=2$ superconformal field theory in four dimensions. 
The complexified coupling for the
diagonal $U(K)$ is identified with the coupling, 
$\tau$, of the original ${\cal N}=4$ theory. The gauge couplings for
the remaining off-diagonal $SU(K)$ factors encode the electric and
magnetic Wilson lines of the DLCQ theory as
$\tau_{i}=(Z_{i+1}-Z_{i})/2 \pi i$ for $i=1,2,\ldots,N$ with the
identification $Z_{N+1}=Z_{1}$.  
\paragraph{}
The resulting ${\cal N}=2$ theory
is precisely the elliptic quiver theory first solved in
\cite{Witten:1997}. The theory has a Coulomb branch of complex dimension 
$KN-N+1$ parametrized by the 
VEVs of the complex scalars in the vector multiplet of $\hat{G}'$. 
The metric on the Coulomb branch is determined by the corresponding
Seiberg-Witten curve $\Sigma$ and meromorphic differential
$\lambda$ for which we will not need explicit forms in the
following. In fact, the considerations of this paper apply to the
Coulomb branch of any ${\cal N}=2$ superconformal theory in
four dimensions and we will review the general features of these
models in the next section.  In particular, 
the four-dimensional Coulomb branch is
a special K\"{a}hler manifold corresponding to the complex structure
moduli space ${\cal M}(\Sigma)$ of the Seiberg-Witten curve. 
As the theory is conformal, the Coulomb 
branch metric is also scale-invariant. 
\paragraph{}
In order to obtain our target space ${\cal M}_{K,N}$ we are
instructed \cite{Kapustin:1998} 
to compactify the four-dimensional ${\cal N}=2$ theory
described above down to three dimensions on a circle of radius 
$R \sim 1/{\cal A}$. As we review in the next section, the
Coulomb branch of the compactified theory acquires additional
dimensions corresponding to electric and magnetic Wilson lines on the
circle. The resulting space is a fibration of the Jacobian torus
${\cal J}(\Sigma)$ over the original Coulomb branch ${\cal M}(\Sigma)$
of the four-dimensional theory. The scale $R$ enters as the inverse
volume of the fibre.    
The total space of the fibration is a hyper-K\"{a}hler manifold of real
dimension $4(KN-N+1)$ which is identified with ${\cal
  M}_{K,N}$. Importantly, although the metric on ${\cal M}_{K,N}$ is
hard to describe in general, it approaches a simple analytic form
known as the {\em semi-flat metric} in a suitable limit where $R$ goes to
infinity \cite{Seiberg:Witten:1996, Gaiotto:Moore:Neitzke:2008}.   
\paragraph{}
In the following we will consider a quantum mechanical $\sigma$-model
with target space ${\cal  M}_{K,N}$ using the ``Coulomb branch''
description of this manifold reviewed above. As the target space is
hyper-K\"{a}hler the corresponding $\sigma$-model has ${\cal N}=(4,4)$
supersymmetry. However, the compact fibre has a fixed volume set by
the scale $R \sim 1/{\cal A}$. Thus the resulting $\sigma$-model
cannot be conformally-invariant. This is consistent with its
interpretation as the DLCQ of the $(2,0)$ theory on a torus of fixed
area ${\cal A}$. However, if we are to obtain a DLCQ description of
the ${\cal N}=4$ theory in the limit ${\cal A}\rightarrow 0$, then it
must be that $SU(1,1|4)$ superconformal invariance emerges in this
limit. In fact, the model in question is part of a large family which can be
obtained by compactifying ${\cal N}=2$ superconformal field theories 
down to three dimensions. Each of these models give rise to a Coulomb
branch which takes the form of a torus fibration over a
scale-invariant special K\"{a}hler manifold. We will show that 
$SU(1,1|4)$ invariance emerges as required in any model of this type.     

\section{Special K\"{a}hler Geometry in Supersymmetric 
Gauge  Theory}
\label{subsection:fieldtheoryreview}

From now on we will consider the general class of models to which the
quantum mechanical $\sigma$-model decribed above belongs. 
We start by considering a generic 
${\cal N}=2$ supersymmetric gauge theory in four dimensions with gauge
group of rank $r$, and give a
 brief review of how the local\footnote{There is a related
  but different notion of ``local special K\"{a}hler geometry'' used
  in supergravity \cite{Crapsetal:1997}. In our case, ``local'' simply
  means that the description makes sense only on some coordinate
  patch.} description of special K\"{a}hler geometry emerges
 in this context, as well as in the compactification of these theories
 on $\mathbb{R}^3\times S^1$.
\paragraph{}
The potential for the scalars in the vector
multiplet has flat directions admitting a moduli space $\M$ of
vacua known as the Coulomb branch where the gauge group is broken down
to its Cartan subgroup $U(1)^{r}$ by the Higgs mechanism. The
low-energy theory thus includes $r$ massless photons\footnote{Here $m$
  and $n$ denote four-dimensional Lorentz indices.} $A_{m}^{I}$, with
field strength $v_{mn}^{I}$ 
for $I=1,2,\ldots,r$, and their ${\cal N}=2$ superpartners.  
In particular there are $r$ massless complex scalar fields $a^I$ 
whose vacuum expectation values provide coordinates on $\M$.
\paragraph{}
The general form of the low-energy effective action on the Coulomb
branch is already highly constrained by supersymmetry  
\cite{Grimm:Sohnius:Wess:1977,Seiberg:1988}. The bosonic part of the
action must be of the form 
\begin{equation} \mathcal{L} = \frac{1}{4\pi}\im
  \tau_{IJ}\partial_m a^I\partial^m\bar{a}^J +
  \frac{1}{8\pi}\im \tau_{IJ}v_{m n}^I v^{J m n} +
  \frac{1}{8\pi}\re \tau_{IJ}
v_{m n}^I\tilde{v}^{J m n}.\label{eqn:4daction}\end{equation}
Here $\tau_{IJ}(a)$ is the matrix of 
complexified gauge couplings, and $\N=2$ supersymmetry forces
\begin{equation} \tau_{IJ} = \frac{\partial^2 \F}{\partial
    a^I \partial a^J}\label{eqn:SKmetric} 
\end{equation}
with $\F(a)$ a holomorphic function known as the prepotential. 
\paragraph{}
As usual, the coefficient matrix of the scalar kinetic terms in 
(\ref{eqn:4daction}) defines a natural metric on $\M$:
\[ ds^2 = \im \tau_{IJ}da^Id\bar{a}^J.\] 
As argued in \cite{Seiberg:Witten:1994}, this cannot be a good global
description because the harmonic function $\im \tau$ is unbounded
below, leading to an indefinite metric. In fact, Seiberg and Witten
\cite{Seiberg:Witten:1994,Seiberg:Witten:1994a}
 were able to give a
global description allowing them to compute the exact quantum
prepotential. The construction relies on an identification of the
Coulomb branch with the moduli space ${\cal M}(\Sigma)$ 
of a certain family of complex
algebraic curves $\Sigma$ of genus $r$, whose period matrices
correspond to the couplings $\tau_{IJ}$. A recent pedagogical
review of these curves is given in \cite{Tachikawa:2013}. 
\paragraph{}
The structure summarised above, namely a complex manifold $\M$ with a
special holomorphic coordinate system $a^I$ such that the metric can
be expressed in terms of a prepotential $\F$ as in
(\ref{eqn:SKmetric}), is known as special K\"{a}hler
geometry. Note in particular that such a 
space is indeed always K\"{a}hler, with potential
\[ K = \im \left(\frac{\partial \F}{\partial a^I}\bar{a}^I\right).\]
\paragraph{}
Now consider compactifying the above class of theories on
$\mathbb{R}^3 \times S^1_R$ as in
\cite{Seiberg:Witten:1996}
. Here $R$ is the radius of $S^1$ and we
will be particularly interested in the case when $R$ is much larger
than any other length scales in the problem. In the context of the
DLCQ model of Section 3, this corresponds to the limit where the size
of the torus
on which the $(2,0)$ theory is defined goes to zero and we are left with 
${\cal N}=4$ supersymmetric Yang-Mills.  
Fortunately it is in this limit where the structure of the
Coulomb branch is simplest and can be understood by 
compactifying the four-dimensional low-energy theory
(\ref{eqn:4daction}). 
In addition to the complex scalars $a^I$ the compactified theory also
contains new real periodic scalars $(\theta_e^I,\theta_{m,I})$
corresponding to the $U(1)^{r}$ electric and magnetic Wilson lines around
$S^1$. These parameterise a complex 
torus $T^{2r}$ which can be identified with the Jacobian
\[ {\cal J}(\Sigma) = 
\frac{\mathbb{C}^r}{\mathbb{Z}^r\oplus \tau \mathbb{Z}^r}\]
of the Seiberg-Witten curve $\Sigma$. Following 
\cite{Gaiotto:Moore:Neitzke:2008} we define 
a complex coordinate
\[ z_I = \theta_{m,I} - \tau_{IJ}\theta^J_e\]
and 1-form\footnote{While this form is closed and equal to $dz_I$ on
  $\mathcal{J}(\Sigma)$, this is no longer true 
 on the full Coulomb branch. We will have more to say about this
in section \ref{section:global}.}
\[\delta z_I = d\theta_{m,I}-\tau_{IJ}d\theta^J_e,\]
in terms of which the metric on ${\cal J}(\Sigma)$ is
\[ ds^2 = \frac{1}{4\pi^2R}\inv^{IJ}\delta z_I\delta \bar{z}_J.\] 
The full Coulomb branch is therefore the total space of 
a fibre bundle\footnote{This statement is not quite precise due to an
  additional subtlety in the global definition of fibre coordinates known as
  the quadratic refinement \cite{Gaiotto:Moore:Neitzke:2008}. 
However, this will not play a role in the 
following.} $\mathcal{B}
\rightarrow \M$ 
over the special K\"{a}hler base $\M$ whose fibres are the Jacobian
tori ${\cal J}$.  When $R$ is much larger than any other scales in the
problem the metric on the total space takes its {\em semi-flat} form: 
\begin{equation} G = R\im \tau_{IJ}da^Id\bar{a}^J +
  \frac{1}{4\pi^2R}\inv^{IJ}\delta z_I\delta 
\bar{z}_J.\label{eqn:Gsf}\end{equation}
As we review below, this is a hyper-K\"{a}hler metric 
as required by supersymmetry. Away from the regime of large $R$, 
the metric is considerably more complicated. In
particular, the metric, which remains hyper-K\"{a}hler for any $R$,
receives instanton corrections of order $\exp(-M_{\rm BPS}R)$ 
where $M_{\rm BPS}$ are the masses of the BPS states of the
four-dimensional theory\footnote{At weak coupling, the lightest
  charged BPS states are the W-bosons which yield corrections of order 
$\exp(-|a|R)$ where $a$ is an integer linear combination
of the scalar VEVs $\langle a^{I} \rangle$.}. These corrections play
an important role in resolving the singularities of the semi-flat
metric. There is a
twistorial approach which yields integral equations determining the
exact metric \cite{Gaiotto:Moore:Neitzke:2008} but we will not need it here.

\section{Superconformal Quantum Mechanics} 
\label{subsection:SCQMreview}

The possible symmetries of quantum mechanical models with generic
curved target spaces are strongly constrained by the presence of
additional geometric structures in the target
\cite{AlvarezGaume:Freedman:1981,Michelson:Strominger:1999,
AS:2014,FOF:Kohl:Spence:1997}. To
make this paper self-contained, we will now 
give a brief review of the key points.
\paragraph{}
We first consider some general features of quantum mechanical 
$\sigma$-models with (at least) ${\cal N}=(1,1)$ supersymmetry. 
Such models contain fermions which satisfy canonical 
anticommutators of the form
\[ \left\{\psi^{\mu},\psi^{\dagger \nu}\right\} = g^{\mu \nu},\]
where $g_{\mu\nu}$ is the target space metric. These 
operators may be used to build up a Fock space in the usual fashion. 
By virtue of Fermi-Dirac
statistics, this Fock space may be 
identified with the exterior algebra of differential forms on the
target \cite{Witten:1982a,Witten:1982b}. States of fermion number zero
are described by ordinary functions (or zero-forms) on the target space. 
There is generically a pair
of supercharges $Q,Q^{\dagger}$ which may be represented via the 
exterior derivative and its adjoint, and the Hamiltonian in this
context is naturally the 
Laplacian acting on forms:
\[ \Delta = dd^{\dagger}+d^{\dagger}d.\]
\paragraph{}
Additional supersymmetries require extra structure on the target space
\cite{AlvarezGaume:Freedman:1981}. $\N=(2,2)$ is obtained if and only
if the metric is K\"{a}hler, and the new supercharges are 
realised by splitting $d$ into the Dolbeault operators
\begin{equation} d = \partial + \bar{\partial}.
\label{eqn:Dolbeault}\end{equation}
The fact that these objects produce the correct supersymmetry algebra
may be taken as a definition of K\"{a}hler geometry
\cite{FOF:Kohl:Spence:1997}. Furthermore the expected $SU(2) \times
U(1)$ R-symmetry emerges naturally from the K\"{a}hler identities and
the associated 
Lefschetz action.
This discussion extends naturally to $\N=(4,4)$ supersymmetry via
hyper-K\"{a}hler geometry
\cite{AlvarezGaume:Freedman:1981,FOF:Kohl:Spence:1997,Verbitsky:1990},
where there is a triplet of 
complex structures $I^a$ and correspondingly three different
decompositions of the exterior derivative as in
(\ref{eqn:Dolbeault}). The corresponding R-symmetry is an $SO(5)$
action generalising 
the Lefschetz action and constructed by Verbitsky in \cite{Verbitsky:1990}.
\paragraph{}
The extension to superconformal invariance also fits into the
geometric framework
\cite{Michelson:Strominger:1999,AS:2014}. Dilatations are generated by
the flow of a vector field $D$ on the target space, hence the
dilatation operator acts as a Lie derivative on the Hilbert space of
differential forms. In order to satisfy the rule
$\left[D,H\right]=2iH$, the coderivative $d^{\dagger} =
(-1)^{np+n+1}*d*$ must be charged under this flow. Since $d$ commutes
with Lie derivatives, 
the solution is that the volume form must expand along the flow, and
hence the vector $D$ must be a homothety, 
satisfying
\[\mathcal{L}_Dg = 2g.\]
The special conformal generator obeys the rules $\left[D,K\right]
=-2iK,~ \left[H,K\right]=-iD$, for which it suffices that $K$ is 
a function on the target space obeying
\begin{equation} \mathcal{L}_DK = 2K, \qquad D_{\mu}
  = \partial_{\mu}K.
\label{eqn:closedhomothety}\end{equation}
A homothety obeying these extra constraints is called closed \cite{Michelson:Strominger:1999}.
\paragraph{}
Adding in $\N=(1,1)$ supersymmetry is straightforward. The
supercharges are as above, and the superconformal charges are defined
via $\left[K,Q\right] = iS$, leading to the 
expressions \cite{AS:2014}
\[ S = idK\wedge, \qquad S^{\dagger} = -ii_D.\]
The closure of the $\left\{Q,S^{\dagger}\right\}$ relations onto the dilatation
is then guaranteed by 
Cartan's formula for the Lie derivative. The resulting model then has
$SU(1,1|1)$ superconformal invariance. 
\paragraph{}
To get extended supersymmetry it is necessary that the homothety
interacts nicely with the complex structure, ensuring that the
Dolbeault supercharges $\partial$ and $\bar{\partial}$ carry the
correct dimensions. It suffices to demand that the
homothety is a holomorphic vector field, $\mathcal{L}_DI = 0$. The
hyper-K\"{a}hler case is similar and requires that $D$ be
triholomorphic. In order that
$\left\{Q,\left[\bar{Q},K\right]\right\}$ closes, it is also necessary
that $K$ is a K\"{a}hler potential \cite{AS:2014}. The above bracket
then produces a K\"{a}hler form, which as already discussed is a
generator for the $SU(2)$ R-symmetry. In the hyper-K\"{a}hler case
the existence of such a potential compatible with all three complex
structures is a non-trivial requirement, but is always met in these
models \cite{AS:2014} (at least assuming the extra 
constraints (\ref{eqn:closedhomothety})) 
thanks to a result of \cite{Hitchinetal:1987}. The resulting models on
K\"{a}hler and hyper-K\"{a}hler manifolds have $U(1,1|2)$ and (a real
form of) $OSp(4|4)$ superconformal invariance respectively. 
\paragraph{}
We will use these results to motivate the superconformal algebra we
introduce in section 
\ref{subsection:su114}, though the final structure will not be
manifestly geometric 
as we will make a truncation to zero fibre momentum. It would be
interesting to have a 
formulation in which the geometry is again made plain.

\section{Hyper-K\"{a}hler Structure of the Coulomb Branch via $T^*\M$} \label{section:global}
Our task is to understand quantum mechanics on the bundle
$\mathcal{B}$ with the 
semi-flat metric (\ref{eqn:Gsf}), for which it will be helpful to
understand the special K\"{a}hler structure in a little more
detail. We use Freed's definition \cite{Freed:1997} of special
K\"{a}hler geometry. The defining feature is the existence of an extra
real torsion-free connection 
$\nabla$ on $T\M$ which is:
\begin{itemize}
\item Flat, $\nabla^2 = 0$
\item Symplectic, $\nabla \omega = 0$
\item ``Special'', $d_{\nabla}I = 0.$
\end{itemize}
Here $\omega$ is the K\"{a}hler form and $I$ the complex
structure. It's important that the final condition is not the same as
$\nabla I=0$. Indeed, if it were then $\nabla$ would be Levi-Civita
and the manifold would be locally isometric to $\mathbb{C}^n$
\cite{Freed:1997}. Rather, the special K\"{a}hler condition may be
written in components as\footnote{View $I$ as a $T\M$-valued 1-form
  and 
act with the exterior covariant derivative $d_{\nabla}$. We label
components of $\nabla$ by 
$\Theta^{\mu}_{\nu \rho}$ to avoid confusion with the Levi-Civita connection.}
\[ \partial_{[\rho}I^{\mu}_{~\nu]} + \Theta^{\mu}_{\sigma [ \rho}I^{\sigma}_{~\nu]} = 0.\]
This connection may be used to establish the existence of local
holomorphic coordinates $a^I$ and prepotential $\F(a)$ satisfying the
characterisation of special K\"{a}hler geometry from section
\ref{subsection:fieldtheoryreview}. Furthermore, being a flat
connection, the only nontrivial consequences of $\nabla$ are
monodromies which turn out to reproduce those of the
Coulomb branch 
from \cite{Seiberg:Witten:1994,Seiberg:Witten:1994a}.
Conversely, a choice of prepotential $\F$ and corresponding special
coordinates $a^I$ is enough to specify a special K\"{a}hler structure
locally 
\cite{Freed:1997}, as we can calculate
\begin{equation} \nabla \frac{\partial}{\partial a^I} =
  -\frac{i}{2}\frac{\partial^3 \F}{\partial a^I \partial a^J \partial
    a^K} da^J \otimes \inv^{KL}\left(\frac{\partial}{\partial a^L} -
    \frac{\partial}{\partial
      \bar{a}^L}\right).\label{eqn:nablacomponents} 
\end{equation}
\paragraph{}
We now turn to the bundle $\mathcal{B}$ and a description of the
semi-flat metric (\ref{eqn:Gsf}). We aim to show that the semi-flat
metric is just the canonical hyper-K\"{a}hler metric on the cotangent
bundle of a special K\"{a}hler manifold as described in
\cite{Freed:1997,Cecotti:Ferrara:Girardello:1988}. A clue about how to
proceed is in the forms $\delta z_I$ used in the semi-flat metric. 
As can be readily checked, it is not true that $\delta z_I$ is the
exterior derivative of $z_I$. 
Instead, we have
\[ dz_I = \delta z_I + \F^{(3)}_{IJK}\inv^{JL}\im z_L da^K.\]
We can make sense of this expression using the theory of horizontal
lifts. Let $X^{\mu}$ be coordinates on some base manifold $M$ and
$P_{\mu}$ the corresponding coordinates on the cotangent bundle
obtained by writing a generic 1-form as $\alpha = P_{\mu}
dX^{\mu}$. Let $\nabla$ be a connection on $TM$ with components
$\Theta^{\mu}_{\nu \rho}$. Then we can define a unique horizontal lift
of the 
frame $\partial_{\mu} \in TM$ to $T(T^*M)$ by setting
\[ D_{\mu} = \frac{\partial}{\partial X^{\mu}} + P_{\rho}
\Theta^{\rho}_{\mu \nu}\frac{\partial}
{\partial P_{\nu}}.\]
In fact, this can be extended to a frame for $T(T^*M)$ by adjoining
the vertical vectors $\partial/\partial P_{\mu}$, and there is
a corresponding dual 
coframe
\[dX^{\mu}, \quad \delta P_{\mu} = dP_{\mu} - P_{\rho}\Theta^{\rho}_{\mu \nu}dX^{\nu}.\]
Carrying out this construction using the special K\"{a}hler connection
$\nabla$ with components determined by (\ref{eqn:nablacomponents}), 
we find the frame
\begin{equation}D_I = \frac{\partial}{\partial a^I} +
  \F^{(3)}_{IKL}\inv^{JL}\im z_J \frac{\partial}{\partial z_K}, \quad
  \frac{\partial}{\partial z_I}\label{eqn:horizontal}
\end{equation}
for $T(T^*\M)$ and coframe
\begin{equation}da^I, \quad \delta z_I = dz_I -
  \F^{(3)}_{IKL}\inv^{JL}\im z_J da^K.\label{eqn:coframelift}
\end{equation}
The key point to notice in this discussion is that the form
$\delta z_I$ defined by horizontal lift exactly coincides
with the one appearing in the semi-flat metric 
(\ref{eqn:Gsf}).
\paragraph{}
Of course we are not done yet since we've not shown that the Coulomb
branch $\mathcal{B}$ has anything to do with $T^*\M$, nor have we
described the hyper-K\"{a}hler structure. To address these issues we
use the results of \cite{Freed:1997,Donagi:Witten:1995}. The moduli
space $\mathcal{B}$ has the structure of an algebraic integrable
system: in particular, it is a holomorphic
symplectic manifold with a fibre bundle structure as described in
Section \ref{subsection:fieldtheoryreview} such that the holomorphic
symplectic form $\eta$ vanishes on restriction to the
fibres. Furthermore, there is a lattice $\Lambda \cong
\mathbb{Z}^r\oplus \tau \mathbb{Z}^r$ (the dual of the electromagnetic
charge lattice) such that the fibres are just
$\mathcal{J} = \mathbb{C}^r/\Lambda$ and are polarised by $\Lambda^*$.
Finally, theorem 3.4 of \cite{Freed:1997} says that such an integrable
system is equivalent to the quotient of the cotangent bundle of a
special K\"{a}hler manifold by a lattice $\Lambda \subset T^*\M$ whose
dual is flat with respect to $\nabla$, such that $\nabla$ has holonomy
in the duality group $Sp(2n;\mathbb{Z})$ defined by $\Lambda^*$. These
are exactly the conditions met by the Coulomb branch of
\cite{Seiberg:Witten:1996} and its associated charge lattice, so we
make the identification
\begin{equation} \mathcal{B} =
  \frac{T^*\M}{\Lambda} \label{eqn:modulispace} 
\end{equation}
with $\M$ the moduli space of the 4d theory.
\paragraph{}
Describing the hyper-K\"{a}hler structure in the large-$R$ limit is
now relatively straightforward. For the metric, we use the argument of
\cite{Freed:1997}: given a complex vector space $W$ with hermitian
metric $g$ and dual $W^*$, there is a canonical `hyper-K\"{a}hler'
metric $G$ on $W\oplus W^*$ 
given by
\[ G(w_1\oplus x_1,w_2 \oplus x_2) = g(w_1,w_2) + g^{-1}(x_1,x_2), \quad w_i \in W, ~x_i \in W^*.\]
In the special K\"{a}hler case this can be globalised, since the
horizontal lift (\ref{eqn:horizontal}) gives an
identification\footnote{Of course, this identification is true for any
  manifold and any connection $\nabla$. The special K\"{a}hler
  condition is needed to verify that the K\"{a}hler forms on $T^*\M$
  are 
closed.}
\[ T(T^*\M) \cong T\M \oplus T^*\M.\]
Since we already have the well-known metric $\im \tau$ for $\M$, we
can simply read off the 
metric on $T^*\M$
\begin{equation} G = \im \tau_{IJ}da^I d\bar{a}^J + \inv^{IJ}\delta
  z_I \delta \bar{z}_J \label{eqn:Gcotangent} 
\end{equation}
where we used $\delta z$ instead of $dz$ as dictated by horizontal
lifting. But this, after some rescalings, is just the semi-flat metric
(\ref{eqn:Gsf}). We've seen that both the full Coulomb branch for the
theory on $\mathbb{R}^3 \times S^1_R$ and its hyper-K\"{a}hler metric
in the large $R$ limit can be constructed canonically from the
cotangent bundle of the four dimensional 
Coulomb branch.
\paragraph{}
This information is enough to construct
the quantum mechanical $\sigma$-model on $\mathcal{B}$, but if we wish to
discuss symmetries then we'll need knowledge of the full
hyper-K\"{a}hler structure. Fortunately, it is equally as
straightforward to read off the K\"{a}hler forms and complex
structures from the cotangent bundle as it is the metric. Our
 presentation has a preferred complex structure
$I^1$ with respect to which $\partial/\partial a^I$ and
$\partial/\partial z_I$ are holomorphic,
and the corresponding K\"{a}hler 
form is
\begin{equation} \omega_1 = \frac{i}{2}\left(\im \tau_{IJ}da^I \wedge
    d\bar{a}^J + \inv^{IJ}\delta z_I \wedge \delta
    \bar{z}_J\right).\label{eqn:omega1} 
\end{equation}
The other K\"{a}hler forms can be read off from the holomorphic
symplectic form 
\begin{equation} \eta = \omega_2 + i\omega_3 = da^I \wedge \delta
  z_I.\label{eqn:holsymplectic}
\end{equation}
\paragraph{}
To close this section, we observe that a K\"{a}hler potential
corresponding to the preferred complex structure is
\begin{equation} K = \im \left(\frac{\partial \F}{\partial
      a^I}\bar{a}^I\right) + 2\inv^{IJ}\im z_I \im
  z_J.\label{eqn:Kahlerpotential} 
\end{equation}
Note however that this is certainly not a hyper-K\"{a}hler potential.
Indeed, it was shown in \cite{Hitchinetal:1987} that such an object
requires the existence of an isometric action of $SU(2)$ (with extra
conditions), which there's no reason to expect our metric
(\ref{eqn:Gcotangent}) to exhibit in general.

\section{Constructing the $\sigma$-Model} \label{section:sigmamodel}
We now have all the necessary ingredients to construct the quantum
mechanics on $\mathcal{B}$. The model fits into the general 
$\N=(1,1)$ form
\begin{equation} S = \int dt \; \frac{1}{2} g_{\mu \nu}\dot{X}^{\mu}
  \dot{X}^{\nu} + i g_{\mu \nu} \psi^{\dagger \mu}
  \frac{D}{dt}\psi^{\nu} + \frac{1}{4}R_{\mu \nu \rho
    \sigma}\psi^{\dagger \mu}
\psi^{\rho}\psi^{\dagger \nu}\psi^{\sigma}\label{eqn:genericaction} \end{equation}
Here $X^{\mu}$ 
are generic target space coordinates and $\psi^{\mu}$ are their
1-complex-component fermionic superpartners. 
The fermion covariant derivative is
\begin{equation} \frac{D}{dt}\psi^{\mu} = \dot{\psi}^{\mu} +
  \dot{X}^{\rho}\Gamma_{\rho \nu}^{\mu}\psi^{\nu}.\label{eqn:Ddt}
\end{equation}
To formulate this model on $\mathcal{B}$ we need the Levi-Civita
connection and curvature associated to the semi-flat metric
(\ref{eqn:Gcotangent}) on $T^*\M$. Explicit expressions
for these are given in appendix 
\ref{appendix:geometry}.
\paragraph{}
A few words on notation are in order at this point. The expression
(\ref{eqn:genericaction}) is of course tensorial, so our convention up
to now of using the indices $I,J,K,\dots$ for everything is no longer
sufficient for bookkeeping purposes: it doesn't distinguish
holomorphic/antiholomorphic nor base/fibre indices. The issue is that
the index $I$ does not represent a tensorial transformation property,
rather a transformation under $Sp(2r;\mathbb{Z})$ duality. The
most mathematically respectable way to deal with this would be to use
a vielbein-like formalism to relate `generic' holomorphic coordinates
to our special coordinates as in 
\cite{Crapsetal:1997}. This will be a little cumbersome for our
purposes, so instead we let $I, \bar{I}$ label (anti)holomorphic base
directions, $I',\bar{I}'$ label (anti)holomorphic fibre directions and
continue to work exclusively with special coordinates. If this is done
carefully then no inconsistencies 
can arise.
\paragraph{}
The boson kinetic terms are easy to read off from the metric:
\begin{equation} \mathcal{L}_{\mbox{Bose}} = \im
  \tau_{I\bar{J}}\dot{a}^I\dot{\bar{a}}^J + 
\inv^{I'\bar{J}'}\frac{\delta z_{I'}}{dt}\frac{\delta
  \bar{z}_{\bar{J}'}}{dt}, \label{eqn:Lbose} 
\end{equation}
where
\[ \frac{\delta z_{I'}}{dt} = \dot{z}_{I'}-\F^{(3)}_{I'JK}\inv^{KL}\im z_L \dot{a}^J\]
reflects the fact that we work in the non-coordinate basis
(\ref{eqn:coframelift}). Turning now to fermions, we will denote
horizontal components by $\chi^I$ and vertical components
$\zeta_{I'}$. 
The covariant time derivatives following from (\ref{eqn:Ddt}) and (\ref{eqn:connection}) are
\begin{subequations} \nonumber \begin{align}\frac{D\chi^I}{dt} &=
    \dot{\chi}^I - \frac{i}{2}
\inv^{IL}\F^{(3)}_{JKL}\dot{a}^K\chi^J \\
&+\frac{i}{2}\bar{\F}^{(3)}_{\bar{L}\bar{M}\bar{N}}\inv^{I\bar{L}}\inv^{J'\bar{M}}\inv^{K'\bar{N}}\frac{\delta
  z_{K'}}{dt}\zeta_{J'} \\ \frac{D\zeta_{I'}}{dt} &= 
\dot{\zeta}_{I'}+\frac{i}{2}\inv^{J'L}\F^{(3)}_{I'KL}\dot{a}^K\zeta_{J'}
\\&+ \frac{i}{2}\F^{(3)}_{I'JL}\inv^{L\bar{K}'}\frac{\delta
  \bar{z}_{\bar{K}'}}{dt}\chi^J. \end{align} 
\end{subequations}
The resulting kinetic terms are quite messy, but can be cleared up
somewhat by making the redefinition
\begin{equation} \zeta^I = \inv^{I\bar{J}'}\zeta_{\bar{J}'} \label{eqn:zetaredef} \end{equation}
and using the base Christoffel symbols
\begin{equation} \Gamma^I_{JK} = -\frac{i}{2}\F^{(3)}_{JKL}\inv^{IL}.\label{eqn:baseChristoffel}\end{equation}
After making these substitutions we find
\begin{eqnarray} 
\mathcal{L}_{\mbox{2-fermi}} &=  & i \im
    \tau_{I\bar{J}}\left[\chi^{\dagger \bar{J}}D_t\chi^I +
      \zeta^{\dagger \bar{J}}D_t\zeta^I\right]  \nonumber \\ &+ &
    i\left[\chi^{\dagger
        \bar{J}}\zeta^{\bar{M}}+\zeta^{\dagger\bar{J}}\chi^{\bar{M}}\right]\im
    \tau_{I\bar{N}}\inv^{K'I}\Gamma^{\bar{N}}_{\bar{J}\bar{M}}\frac{\delta
      z_{K'}}{dt} + \mbox{ conjugates} 
\label{eqn:L2fermi}  
\end{eqnarray}
where
\[ D_t\chi^I = \dot{\chi}^I +\Gamma^I_{JK}\dot{a}^J\chi^K\]
is the base space covariant derivative. We note in passing that $\chi$
and $\zeta$ appear symmetrically in this expression, which suggests
the possibility of combining them into a single object. In fact this
will be exactly what we do in Section \ref{subsection:SO6} to
exhibit the enhancement of R-symmetry from the $SO(5)$ present in any
 hyper-K\"{a}hler model to the $SO(6) \subset SU(1,1|4)$ required by DLCQ.
\paragraph{}
As may be seen from the form of the curvature components
(\ref{eqn:curvature}), the four-fermion terms fall into two broad
classes: contractions with the base space Riemann tensor
\begin{equation}R_{I\bar{J}K\bar{L}} = -\frac{1}{4}\inv^{M\bar{N}}\F^{(3)}_{IKM}\bar{\F}^{(3)}_{\bar{J}\bar{L}\bar{N}}
\label{eqn:baseRiemann} \end{equation}
and contractions with the totally symmetric base space tensor
\begin{align} \begin{aligned} G_{IJKL} &= -\frac{i}{2}\nabla_I \F^{(3)}_{JKL}
 \\&= -\frac{i}{2}\F^{(4)}_{IJKL} +
 \frac{1}{4} \inv^{MN}\left(\F^{(3)}_{ILM}\F^{(3)}_{JKN}
 + \F^{(3)}_{JLM}\F^{(3)}_{IKN} + 
\F^{(3)}_{KLM}\F^{(3)}_{IJN}\right). \end{aligned}
 \label{eqn:chiraltensor}\end{align}
After using the
same redefinition of $\zeta$ as for the two-fermion terms
(\ref{eqn:zetaredef}), the latter 
type gives
\begin{equation} \mathcal{L}_{\mbox{4-fermi (a)}} = 2\re
  \left[G_{IJKL}\chi^{\dagger I}\chi^J \zeta^{\dagger
      K}\zeta^L\right]. \label{eqn:L4fermichiral} 
\end{equation}
The Riemann tensor terms are somewhat messier: we find
\begin{eqnarray} \label{eqn:L4fermiRiemann} 
    \mathcal{L}_{\mbox{4-fermi (b)}} & = &
    R_{I\bar{J}K\bar{L}}\left[\chi^{\dagger I} \chi^K \chi^{\dagger
        \bar{J}} \chi^{\bar{L}} + \chi^{\dagger I}\zeta^K\chi^{\dagger
        \bar{J}}\zeta^{\bar{L}} + \zeta^{\dagger
        I}\chi^{K}\zeta^{\dagger \bar{J}} \chi^{\bar{L}}\right. \nonumber \\
 & & \left.+ \zeta^{\dagger I}\zeta^K\zeta^{\dagger \bar{J}}
      \zeta^{\bar{L}} + \zeta^{\dagger I}\chi^{\dagger
        K}\chi^{\bar{J}}\zeta^{\bar{L}} + \chi^I \zeta^K \chi^{\dagger
        \bar{J}}\chi^{\dagger \bar{L}}\right]  
\end{eqnarray}
Although these terms are not especially enlightening
at the moment, we will see in Section \ref{subsection:SO6} that they
come in exactly the right combinations to admit an extension to
$SO(6)$ R-symmetry. The full Lagrangian is the sum of 
(\ref{eqn:Lbose}), (\ref{eqn:L2fermi}), (\ref{eqn:L4fermichiral}) 
and (\ref{eqn:L4fermiRiemann}). 
\paragraph{}
In identifying the symmetries of our model it
will be most convenient to work in the Hamiltonian formalism. This is
essentially because the symmetry properties of the objects 
$\delta z/dt$ are somewhat mysterious and will become much
clearer after Legendre transform. To that end, we begin 
by computing the canonical momenta
\begin{equation} \label{eqn:momenta} \begin{aligned} P_I =
    \frac{\partial \mathcal{L}}{\partial \dot{a}^I} &= \im
    \tau_{I\bar{J}}\dot{\bar{a}}^{\bar{J}} + 2i\im z_K \frac{\delta
      \bar{z}_{\bar{J}'}}{dt}\frac{\partial}{\partial a^I}
    \inv^{\bar{J}'K} \\ &+ \frac{1}{2} \im
    \tau_{K\bar{J}}\inv^{KM}\F^{(3)}_{ILM}\left(\chi^{\dagger
        \bar{J}}\chi^L + \zeta^{\dagger \bar{J}} \zeta^L\right) \\
    &- 2R_{I\bar{J}K\bar{L}}\inv^{KM}\im z_M
    \left(\chi^{\dagger \bar{L}} \zeta^{\bar{J}} + \zeta^{\dagger
        \bar{M}}\chi^{\bar{J}}\right) \\ P^{I'} = \frac{\partial
      \mathcal{L}}{\partial \dot{z}_{I'}} & = \inv^{I'\bar{J}'}\frac{\delta \bar{z}_{\bar{J}'}}{dt} \\
    &-\frac{1}{2}\inv^{I'\bar{L}}\bar{\F}^{(3)}_{\bar{J}\bar{K}\bar{L}}\left(\chi^{\dagger
        \bar{K}}\zeta^{\bar{J}}+\zeta^{\dagger
        \bar{K}}\chi^{\bar{J}}\right). \end{aligned}
\end{equation}
Following
\cite{Michelson:Strominger:1999,AS:2014} we also define the 
covariant momenta
\begin{equation} \label{eqn:covariantmomenta} \Pi_I
    = \im \tau_{I\bar{J}}\dot{\bar{a}}^{\bar{J}}, \qquad
    \Pi^{I'} = \inv^{I'\bar{J}'}\frac{\delta \bar{z}_{\bar{J}'}}{dt}
\end{equation}
in terms of which the Hamiltonian is simply
\begin{equation} H = \inv^{I\bar{J}}\Pi_I\bar{\Pi}_{\bar{J}} + \im
  \tau_{I'\bar{J}'}\Pi^{I'}\bar{\Pi}^{\bar{J}'} -
  \mathcal{L}_{\mbox{4-fermi}}. \label{eqn:Hamiltonian} 
\end{equation}
\paragraph{}
The commutation relations of these objects are subtle and
require the Dirac bracket procedure to get right, the details
 of which we omit. We obtain the non-vanishing commutators:
\begin{subequations} \nonumber \label{eqn:Diracbrackets} \begin{align}
    \left[a^I,\Pi_J\right] &= i\delta^I_J\\ \left[z_{I'},\Pi_J\right]
    &= i\im z_K \F^{(3)}_{I'JL}\inv^{KL} & \left[z_{I'},P^{J'}\right]
    &= i\delta_{I'}^{J'}  \\ \left[\Pi_I,P^{J'}\right] &=
    \frac{1}{2}\F^{(3)}_{IK'L}\inv^{J'L}P^{K'} &
    \left[\Pi_I,P^{\bar{J}'}\right] &=
    -\frac{1}{2}\F^{(3)}_{IK'L}\inv^{\bar{J}'L}P^{K'} \\
    \left\{\chi^I,\chi^{\dagger \bar{J}}\right\} &= \inv^{I\bar{J}} &
    \left\{\zeta^I,\zeta^{\dagger \bar{J}}\right\} &= \inv^{I\bar{J}}
    \\ \left[\Pi_I,\chi^J\right] &= i\Gamma^J_{IK}\chi^K &
    \left[\Pi_I,\chi^{\dagger J}\right] &= i\Gamma^J_{IK}\chi^{\dagger
      K} \\ \left[\Pi_I,\zeta^J\right] &= i\Gamma^J_{IK} \zeta^K &
    \left[\Pi_I,\zeta^{\dagger J}\right] &=
    i\Gamma^J_{IK}\zeta^{\dagger K} \end{align} 
\end{subequations}
It is important to notice that
the commutation relations of $P^{I'}$ are consistent with zero, so we
can truncate to the sector of zero momentum around the fibres. Of
course such a truncation is the natural one to consider in the $R
\rightarrow \infty$ limit of the moduli space, in which the torus
fibres become small \cite{Seiberg:Witten:1996}. In the following
sections we will see that this is crucial to revealing the 
superconformal symmetry of our model.

\section{R-Symmetry Enhancement $SO(5) \rightarrow SO(6)$} \label{subsection:SO6}
We can now begin our analysis of the symmetries of our model. As
 reviewed in Section \ref{subsection:SCQMreview}, a
hyper-K\"{a}hler $\sigma$-model must have $\N = (4,4)$ supersymmetry
 with $SO(5)$ R-symmetry acting
purely on fermions. In fact we will see that in this case
the R-symmetry extends to $SO(6)$, but it will be a good first step to
put the Hamiltonian into manifestly $SO(5)$-invariant form and
construct the generators and 
supersymmetries.\footnote{We use $A,B$ to index the $\bf{4}$ of SO(6) and
 $\bar{A},\bar{B}$ the $\bar{\bf{4}}$. Indices are raised/lowered with
 $\delta_{A\bar{B}} = \mbox{diag}(1,1,1,1)$. We keep the $\bf{4}$ and
 $\bar{\bf{4}}$ separate even in $SO(5)$, in view of the forthcoming extension
to $SO(6)$. $SO(5)$ has antisymmetric invariant tensor $\Omega_{AB}$ and where necessary we take $\Omega_{23}=\Omega_{41}=1$.}
\paragraph{}
The $SO(5)$ R-symmetry
generators are as given in \cite{AS:2014}: in a notation emphasising
the $SU(2)$ subgroups associated to 
each complex structure $I^a$ they are
\begin{align} \label{eqn:SO5} \begin{aligned} J_+^a &=
    \frac{1}{2}\omega^a_{\mu \nu}
\psi^{\dagger \mu}\psi^{\dagger \nu} \\
 R^a &=
-\frac{i}{2}\omega^a_{\mu \nu} \psi^{\dagger \mu}\psi^{\nu} \end{aligned}
&&
\begin{aligned} J_-^a &=
\frac{1}{2}\omega^a_{\mu \nu} \psi^{\nu}\psi^{\mu} \\ J_3 &=
\frac{1}{2}\left(g_{\mu \nu}\psi^{\dagger \mu}\psi^{\nu} -
  2r\right). \end{aligned} \end{align} 
In terms of the canonical quantisation in which wavefunctions with
fermion number $F$ become differential forms of degree $F$, $J^a_+$ is
wedging with the K\"{a}hler form $\omega^a$, $R^a$ is the action of
$I^a$ and $J_3$ counts degree.
We can easily read off explicit expressions for these generators from
(\ref{eqn:omega1}) and (\ref{eqn:holsymplectic}) using the rules
\[ da^I \leftrightarrow \chi^{\dagger I}, \quad \inv^{J\bar{I}'}
\delta \bar{z}_{\bar{I}'}\leftrightarrow \zeta^{\dagger J}.\]
The detailed coefficients are not important, but notice that all generators follow the pattern
\[ T \sim \im \tau \times \mbox{holomorphic fermion} \times \mbox{antiholomorphic fermion}.\]
This means that the (anti)holomorphic fermions carry separate actions
of $SO(5)$. Indeed, if we define the objects
\[ \psi^{IA} = \left(\chi^I,\chi^{\dagger I},\zeta^I,\zeta^{\dagger
    I}\right)\]
then we see that $\psi^{IA}$ transforms in the $\mathbf{4}$ and
$\bar{\psi}^{\bar{I}\bar{A}} = \left(\psi^{IA}\right)^{\dagger}$ 
in the $\mathbf{\bar{4}}$. They satisfy the simple anticommutation relations
\begin{equation} \nonumber
\left\{\psi^{IA},\bar{\psi}^{\bar{J}\bar{B}}\right\} =
    \delta^{A\bar{B}}\inv^{I\bar{J}}.
\end{equation}
\paragraph{}
We now put the Hamiltonian (\ref{eqn:Hamiltonian}) into a
manifestly $SO(5)$-invariant form. 
To begin with note that we have
\[ \left[\Pi_I,\psi^{JA}\right] = i\Gamma^J_{IK}\psi^{KA}, \quad
\left[\Pi_I,\bar{\psi}^{\bar{J}\bar{A}}\right] = 
0\]
so that $\Pi_I$ must be $SO(5)$-neutral and the term
$\inv^{I\bar{J}}\Pi_I\bar{\Pi}_{\bar{J}}$ in the Hamiltonian is
$SO(5)$-invariant. Another straightforward part is the chiral 
term
\[ 2\re \left(G_{IJKL}\chi^{\dagger I}\chi^J\zeta^{\dagger K}\zeta^L\right)\]
which may be written as
\begin{equation} H_{\mbox{chiral}} =  \frac{1}{12} \re
  \left(\epsilon_{ABCD}G_{IJKL}\psi^{IA}\psi^{JB}\psi^{KC}\psi^{LD}\right)\label{eqn:Hchiral}
\end{equation}
using the symmetry of $G_{IJKL}$.
\paragraph{}
The remaining terms are less obvious. It will prove convenient to work
in terms of the canonical momentum $P^{I'}$ rather than its covariant
form $\Pi^{I'}$, 
so that
\begin{eqnarray} \label{eqn:PisquaredinH} 
\im
    \tau_{I'\bar{J}'}\Pi^{I'}\bar{\Pi}^{\bar{J}'} &= & \im \tau_{I'
      \bar{J}'}P^{I'}\bar{P}^{\bar{J}'} + \re
    \left[\F^{(3)}_{I'JK}\left(\chi^{\dagger J} \zeta^K +
        \zeta^{\dagger J}\chi^K\right)P^{I'}\right] \nonumber \\ 
&+& R_{I\bar{J}K\bar{L}}\left(\chi^{\dagger I} \zeta^K +
      \zeta^{\dagger I} \chi^K\right)\left(\chi^{\dagger \bar{J}}
      \zeta^{\bar{L}} + \zeta^{\dagger \bar{J}}
      \chi^{\bar{L}}\right). 
\end{eqnarray}
The first term on the right is manifestly 
$SO(5)$-invariant and the second can be put in the form
\[ \re \left[\F^{(3)}_{I'JK}\left(\chi^{\dagger J} \zeta^K +
    \zeta^{\dagger J}\chi^K\right)P^{I'}\right] = \frac{1}{2}\re
\left(\F^{(3)}_{I'JK}\Omega_{AB}
\psi^{JA}\psi^{KB}P^{I'}\right).\]
We emphasise for later the use of the $SO(5)$ symplectic form
$\Omega_{AB}$ which will clearly obstruct any possible extension
to $SO(6)$. The four-fermion terms in (\ref{eqn:PisquaredinH}) can be
combined with the remainder of the Hamiltonian
(\ref{eqn:L4fermiRiemann}) to obtain
\begin{equation} \label{eqn:HRiemann} H_{\mbox{Riemann}} =
  \frac{1}{2}R_{I\bar{J}K\bar{L}}\psi^{IA}\bar{\psi}^{\bar{J}}_A
  \psi^{KB}\bar{\psi}^{\bar{L}}_B. 
\end{equation}
Taking everything together, we have the $SO(5)$ invariant Hamiltonian
\begin{eqnarray} \label{eqn:invariantHamiltonian}  H &= & 
    \inv^{I\bar{J}}\Pi_I\bar{\Pi}_{\bar{J}} + \frac{1}{12} \re
    \left(\epsilon_{ABCD}G_{IJKL}\psi^{IA}\psi^{JB}\psi^{KC}\psi^{LD}\right)
    \nonumber \\ &+ & 
    \frac{1}{2}R_{I\bar{J}K\bar{L}}\psi^{IA}\bar{\psi}^{\bar{J}}_A
    \psi^{KB}\bar{\psi}^{\bar{L}}_B \nonumber \\ &+  & \im
    \tau_{I'\bar{J}'}P^{I'}P^{\bar{J}'} + \frac{1}{2}\re
    \left(\F^{(3)}_{I'JK}\Omega_{AB}\psi^{JA}\psi^{KB}P^{I'}\right) 
\end{eqnarray}
\paragraph{}
We can also put the supercharges into $SO(5)$ multiplets. In a generic
hyper-K\"{a}hler $\sigma$-model of the form (\ref{eqn:genericaction})
these charges are (see e.g 
\cite{AS:2014})
\begin{align} \nonumber Q &= i\psi^{\dagger
      \mu}\Pi_{\mu} & Q^a &=
    -i\psi^{\dagger \mu} I^{a\nu}_{\mu}\Pi_{\nu}\end{align}
where we were not careful about operator ordering\footnote{In
  \cite{AS:2014} we used $Q^{\dagger}=-i\Pi_{\mu}\psi^{\mu}$ in order
  to ensure the validity of the exterior algebra representation $Q
  \rightarrow d,Q^{\dagger} \rightarrow d^{\dagger}$. In this paper,
  we use a different ordering convention to make the $SO(5)$
  invariance manifest. Strictly speaking, much of what follows is
 only valid at the level of Poisson brackets, but we do not
 anticipate this causing any problems}. Using the complex structures
(\ref{eqn:holsymplectic}) we can easily read off the charges:
\begin{equation} \nonumber Q = i\chi^{\dagger
      I}\Pi_I + i \im \tau_{I'\bar{J}}\zeta^{\dagger \bar{J}}P^{I'} +
    \frac{i}{2}\zeta^{\dagger L}\F^{(3)}_{JLM}\left(\chi^{\dagger
        M}\zeta^J + \zeta^{\dagger M}\chi^J\right) - \mbox{ complex
      conjugate}, \end{equation}
with similar expressions for $Q^a$.
Taking suitable linear combinations of these leads to expressions which
manifestly transform in the $\mathbf{4}$ of $SO(5)$,
\begin{equation} \label{eqn:SUSYcharges} Q^A = \psi^{IA}\Pi_I +
  \frac{1}{12}\epsilon^A_{~\bar{B}\bar{C}\bar{D}}\bar{\F}^{(3)}_{\bar{I}\bar{J}\bar{K}}
\bar{\psi}^{\bar{I}\bar{B}}\bar{\psi}^{\bar{J}\bar{C}}
\bar{\psi}^{\bar{K}\bar{D}}
  + \im \tau_{I'\bar{J}}P^{I'}\Omega^A_{~\bar{B}}\bar{\psi}^{\bar{J}\bar{B}} 
\end{equation}
along with the conjugate $\bar{Q}^{\bar{A}} =
\left(Q^A\right)^{\dagger}$ which transforms in the
$\mathbf{\bar{4}}$. These
charges obey the standard supersymmetry algebra
\begin{subequations} \nonumber \begin{align} \left\{Q^A,Q^B\right\} &=
    0\\
    \left\{Q^A,\bar{Q}^{\bar{B}}\right\} &=
    \delta^{A\bar{B}}H. \end{align} 
\end{subequations}
\paragraph{}
As remarked briefly above, it is clear from the $SO(5)$-manifest form
of both the Hamiltonian and the supercharges that $SO(5)$ is the
largest symmetry we can get without changing something, since
the expressions (\ref{eqn:invariantHamiltonian}) and
(\ref{eqn:SUSYcharges}) both require the $SO(5)$-invariant tensor
$\Omega_{AB}$ which does not exist in $SO(6)$. Furthermore we do not
expect conformal invariance without some modification, as the torus 
fibre has a fixed finite size. In the following we tackle each
 extension in turn, and show that they can
both be achieved via the same truncation to the
sector of zero fibre momentum.
\paragraph{}
We can extend the $SO(5)$ R-symmetry (\ref{eqn:SO5}) to $SO(6) \simeq SU(4)$
via the obvious generalisation
\begin{equation} R^{A\bar{B}} = i\im \tau_{I\bar{J}}\left(\psi^{IA}\bar{\psi}^{\bar{J}\bar{B}}
-\frac{1}{4}\delta^{A\bar{B}}\psi^{IC}\bar{\psi}^{\bar{J}}_C\right), \label{eqn:SO6} \end{equation}
where the second term removes a trace part 
and reduces $U(4) \rightarrow SU(4)$. These obey
 the expected commutation relations
\begin{subequations} \nonumber \begin{align}
    \left[R^{A\bar{B}},R^{C\bar{D}}\right] &=
    i\left(\delta^{C\bar{B}}R^{A\bar{D}} -
      \delta^{A\bar{D}}R^{C\bar{B}}\right) \\
    \left[R^{A\bar{B}},\psi^{IC}\right] &=
    i\left(\delta^{C\bar{B}}\psi^{IA} -
      \frac{1}{4}\delta^{A\bar{B}}\psi^{IC}\right) \\
    \left[R^{A\bar{B}},\bar{\psi}^{\bar{I}\bar{C}}\right] &=
    -i\left(\delta^{A\bar{C}}\bar{\psi}^{\bar{I}\bar{B}}-\frac{1}{4}\delta^{A\bar{B}}
\bar{\psi}^{\bar{I}\bar{C}}\right)
    \\ \left[R^{A\bar{B}},\Pi_I\right] &= 0 =
    \left[R^{A\bar{B}},P^{I'}\right],
\end{align} \end{subequations}
which confirm that $\psi$ transforms in the $\mathbf{4}$, $\bar{\psi}$
in the $\mathbf{\bar{4}}$ 
and that $\Pi_I,P^{I'}$ are neutral.
\paragraph{}
This is enough to demonstrate that the majority of the terms in both
the Hamiltonian (\ref{eqn:invariantHamiltonian}) and the supercharges
(\ref{eqn:SUSYcharges}) have the correct charges under $SO(6)$. The
problematic terms are of course those relying on the $SO(5)$-invariant
form $\Omega_{AB}$, but we note that they always appear
multiplying the fibre momentum $P^{I'}$. Recall that we are
considering the compactification on $\mathbb{R}^3 \times S^1_R$ in the
limit $R\rightarrow \infty$ where the torus fibres $\cal{J}$ become
small. It is then natural to truncate to zero fibre momentum,
since we can Fourier expand around the
 fibres\footnote{This is true at least on a locally trivial patch of the
  fibre bundle.}
and see that states with nonzero $P^{I'}$ have divergent energy. We
conclude that our model admits $SO(6)$ R-symmetry at large $R$ as
required by DLCQ.

\section{$SU(1,1|4)$ and Scale-Invariant Special K\"{a}hler
  Geometry} 
\label{subsection:su114}
Finally we turn to superconformal invariance. Recall that
$SU(1,1|4)$ is a simple supergroup with 
bosonic part
\[ SO(2,1)\times U(4)\]
and a total of 16 fermions: $\left(Q^A,S^B\right)$ transform in the
$(\mathbf{2},\mathbf{4})$ and
$\left(\bar{Q}^{\bar{A}},\bar{S}^{\bar{B}}\right)$ in the 
$(\mathbf{2},\mathbf{\bar{4}})$.
\paragraph{}
As reviewed in Section \ref{subsection:SCQMreview}, superconformal
invariance requires the target space to admit a homothety, that is a
vector $D$ satisfying $\mathcal{L}_Dg = 2g$, which acts as the
dilatation operator. There is no reason why a generic special
K\"{a}hler manifold might be expected to admit such an object, so in
order to proceed we need to make a definition. We call a geometry
\textit{scale-invariant special K\"{a}hler} (SISK) if there is a prepotential
satisfying the further condition 
\begin{equation} a^I\frac{\partial}{\partial a^I} \F = 2\F. \label{eqn:conformalSK} \end{equation}
If the prepotential is of this form then it is clear that
the Coulomb branch of the 4d theory has a homothety
\begin{equation} D = a^I\frac{\partial}{\partial a^I} +
  \bar{a}^I\frac{\partial}{\partial \bar{a}^I}.\label{eqn:homothety} 
\end{equation}
\paragraph{}
One might wonder whether the SISK condition has any interesting
solutions. A trivial one has $\F$ a quadratic polynomial in the $a^I$,
corresponding to a flat manifold. This is of some limited
physical interest as of course it corresponds to the finite $\N=4$ theory
and to the diagonal $U(1)$ in our quiver model (\ref{eqn:quivergroup}),
but we'd like to do better. The SISK condition follows if and only if $\F$
is homogeneous of degree 2, so any function of the 
form
\[ \F = \left(a^1\right)^2f\left(\frac{a^I}{a^J}\right)\]
for arbitrary holomorphic $f$ will do. There is a large family of such
prepotentials available in physics. It is perhaps no surprise that
they arise from Coulomb branches of $\N=2$ superconformal theories
in four dimensions, whose microscopic scale invariance is reflected
in a scale-invariant metric in the low-energy theory. In particular,
of course, our quiver model of DLCQ is of this form.
\paragraph{}
Returning to the construction of $SU(1,1|4)$, we try to
give an expression for the special conformal generator $K$. As reviewed
 in Section \ref{subsection:SCQMreview},
 this must be given by the K\"{a}hler
potential, but the question here is which one?
(\ref{eqn:Kahlerpotential}) is a possible potential on the total space
$\mathcal{B}$ but only with respect to the preferred complex
structure, whereas \cite{AS:2014} suggests it must be a
hyper-K\"{a}hler potential. Fortunately the same truncation to
$z$-independent functions used in Section \ref{subsection:SO6} comes
to the rescue, and it will turn out to be sufficient to use the base
space K\"{a}hler potential
\begin{equation} K = \im \left(\frac{\partial \F}{\partial a^I}
    \bar{a}^I\right) \label{eqn:specialconformal}
\end{equation}
obtained from (\ref{eqn:Kahlerpotential}) by setting $z = 0$.
\paragraph{}
With special conformal generator as above, it is straightforward to
calculate the dilatation operator using the rule $\left[H,K\right] =
-iD$, and we find (note that we always assume the 
SISK condition from now on)
\begin{equation} D = a^I\Pi_I +
  \bar{a}^{\bar{I}}\bar{\Pi}_{\bar{I}} \label{eqn:dilatation} 
\end{equation}
in agreement with the homothety (\ref{eqn:homothety}). Notice in
particular that the bosons $a^I,\bar{a}^{\bar{I}}$ have dimension 1
\[ \left[D,a^I\right] = -ia^I\]
while, as a consequence of the SISK condition, all fermions have dimension
0. In fact after we make the truncation to $P^{I'} = 0$, 
so that
\begin{eqnarray} \label{eqn:truncatedHamiltonian}  H &= & 
    \inv^{I\bar{J}}\Pi_I\bar{\Pi}_{\bar{J}} + \frac{1}{12} \re
    \left(\epsilon_{ABCD}G_{IJKL}\psi^{IA}\psi^{JB}\psi^{KC}\psi^{LD}\right)
    \nonumber \\ &+ &
    \frac{1}{2}R_{I\bar{J}K\bar{L}}\psi^{IA}\bar{\psi}^{\bar{J}}_A
    \psi^{KB}\bar{\psi}^{\bar{L}}_B,  
\end{eqnarray}
we find that the full $SO(2,1)$ conformal algebra
\[ \left[D,H\right]=2iH, \quad \left[D,K\right]=-2iK,\quad \left[H,K\right]=-iD\]
is obeyed. It is also manifest that $SO(2,1)$ commutes with the
$SO(6)$ R-symmetry (\ref{eqn:SO6}). The extra terms present in $H$ for
$P^{I'}\neq 0$ break the relation $\left[D,H\right]=2iH$ as $P^{I'}$ has
 the wrong dimension. Explicit expressions for the `deformed'
algebra occurring for $P^{I'} \neq 0$ are given in appendix
\ref{appendix:deformation}.
\paragraph{}
Making the same truncation for the supercharges, so they read 
\begin{equation} Q^A = \psi^{IA}\Pi_I +
  \frac{1}{12}\epsilon^A_{~\bar{B}\bar{C}\bar{D}}\bar{\F}^{(3)}_{\bar{I}\bar{J}\bar{K}}
\bar{\psi}^{\bar{I}\bar{B}}\bar{\psi}^{\bar{J}\bar{C}}
\bar{\psi}^{\bar{K}\bar{D}}, 
\label{eqn:truncatedSUSY}
\end{equation}
we find
\[ \left[D,Q^A\right] = iQ^A\]
as required. Again the extra $P^{I'} \neq 0$ terms in
(\ref{eqn:SUSYcharges}) have the wrong dimension. We can also
define the superconformal generators $S$ by the rule
$\left[K,Q^A\right] = iS^A$, 
giving
\begin{equation} S^A = \im
  \tau_{I\bar{J}}\bar{a}^{\bar{J}}\psi^{IA} \label{eqn:superconformal} 
\end{equation}
along with their conjugates $\bar{S}^{\bar{A}}$. These generators have
the correct $SO(6)$ 
transformation properties and dimensions, as well as obeying the expected relations
\begin{align} \nonumber \left\{S^A,S^B\right\} &=
    0 & \left[K,S^A\right] &=0 \\ \nonumber
    \left\{S^A,\bar{S}^{\bar{B}}\right\} &= \delta^{A\bar{B}}K &
    \left[H,S^A\right] &= -iQ^A, \end{align} 
the last relation also being broken by the $P^{I'} \neq 0$ terms in $H$.
\paragraph{}
It remains to check the $\left\{Q,S\right\}$ relations. Doing so reveals a $U(1)$ R-symmetry
\begin{equation} \mathcal{R} = i\left(a^I\Pi_I -
    \bar{a}^{\bar{I}}\bar{\Pi}_{\bar{I}}\right) + \frac{1}{2}\im
  \tau_{I\bar{J}}\psi^{IA}\bar{\psi}^{\bar{J}}_A 
\label{eqn:U1R} \end{equation}
with charges
\[ \begin{array}{cccccccc}  a^I & \bar{a}^{\bar{I}} & \psi^{IA} &
  \bar{\psi}^{\bar{I}\bar{A}} & Q^A & S^A & \bar{Q}^{\bar{A}} &
  \bar{S}^{\bar{A}}\\  1 & -1 & \frac{1}{2} &
  -\frac{1}{2} & -\frac{1}{2} & -\frac{1}{2} & \frac{1}{2} &
  \frac{1}{2} 
\end{array} \]
With this definition we have
\begin{subequations} \nonumber \begin{align} \left\{Q^A,S^B\right\} &=
    0\\
    \left\{Q^A,\bar{S}^{\bar{B}}\right\} &=
    \frac{1}{2}\delta^{A\bar{B}}\left(D-i\mathcal{R}\right) - R^{A\bar{B}},
\end{align} \end{subequations}
where the first relation is also broken when $P^{I'}\neq
0$. This completes our construction of $SU(1,1|4)$: for
convenience, we collect the operators and commutation relations 
in appendix \ref{appendix:su114}.

\section{Reduction from Four Dimensions}
In this section we would like to present a simpler perspective on the
results given above. In the limit where we restrict to states with
zero momentum along the fibre, the model reduces to a $\sigma$-model
with the special K\"{a}hler base as target. In fact  
(\ref{eqn:truncatedHamiltonian}) is the Hamiltonian for a novel type of 
quantum mechanical $\sigma$-model with special  K\"{a}hler target.  
It can also be thought of as quantum
mechanics on the Coulomb branch of a 
four-dimensional gauge theory
with ${\cal N}=2$ superconformal symmetry. From this point of view, it
is natural to suspect that we could have derived it directly from the
low-energy effective action of the 4d theory by dimensional reduction
to quantum mechanics. We will now perform the reduction and discuss
the symmetries of the resulting $\sigma$-model in this context. 
\paragraph{}
We will begin by considering an arbitrary ${\cal N}=2$ supersymmetric
field theory in four dimensions with gauge group of rank $r$.  
The bosonic symmetry of the model includes the Lorentz group
$SO(3,1)\simeq SL(2)_{A}\times SL(2)_{B}$ as well as an $SU(2)$
R-symmetry. If the theory is superconformal, it will also have a
non-anomalous $U(1)$ R-symmetry. The supercharges $Q^{i}_{\alpha}$ and
$\bar{Q}_{\dot{\alpha}i}$
transform in the $({\bf 2},{\bf 1}, {\bf 2})_{+1}\oplus 
({\bf 1},{\bf 2}, {\bf 2})_{-1}$ of the global symmetry group 
\begin{eqnarray}
G_{\rm global} & = & SL(2)_{A}\times SL(2)_{B}\times SU(2)_{R} \times
U(1)_{R}. \nonumber 
\end{eqnarray}
\paragraph{}
The theory has a Coulomb branch where the scalars in the vector
multiplet acquire non-zero expectation values and the gauge group is
broken to $U(1)^{r}$ by the adjoint Higgs mechanism. The massless
fields consist of $r$ $U(1)$ vector multiplets with complex scalars
$a^{I}$ as lowest components, where $I=1,2,\ldots,r$
labels the Cartan subalgebra of the gauge group. The supersymmetry
multiplet combines $a^{I}$ with left-handed Weyl fermions
$\lambda_{\alpha}^{I}$, $\psi^{I}_{\alpha}$ and the self-dual part of
the $U(1)$ gauge field strength, $(v^{\rm SD})^{I}_{mn}$. The charge
conjugate multiplet has lowest component $\bar{a}^{I}$ 
and also includes right-handed Weyl fermions
$\bar{\lambda}_{\dot{\alpha}}^{I}$, $\bar{\psi}^{I}_{\dot{\alpha}}$ 
together with the anti-self-dual part of
the $U(1)$ gauge field strength, $(v^{\rm ASD})^{I}_{mn}$. 
\paragraph{}
The scalars $a^{I}$ and $\bar{a}^{I}$ parameterise a vacuum moduli
space which is a special K\"{a}hler manifold of complex dimension
$r$. The metric is determined in terms of the holomorphic prepotential
${\cal F}(a)$ by (\ref{eqn:SKmetric}).  
We also use the Christoffel symbols (\ref{eqn:baseChristoffel}) to
define a covariant derivative for the fermions: 
\begin{eqnarray}
D_{m}\psi^{I}_{\alpha} & = & \partial_{m}\psi^{I}_{\alpha} \,+\, 
\Gamma^{I}_{JK}\partial_{m}a^{J}\psi^{K}_{\alpha}.
\nonumber 
\end{eqnarray}
\paragraph{}
The full low-energy effective Lagrangian is
 (see e.g \cite{Dorey:Khoze:Mattis:1996})
\begin{eqnarray} 
{\cal L}_{\rm eff} & = & {\cal L}_{1}+ {\cal L}_{2}+{\cal L}_{3}, 
\nonumber 
\end{eqnarray}
where (up to an irrelevant overall factor of $1/4\pi$)
\begin{eqnarray}
{\cal L}_{1} & = & -\im \tau_{IJ}\left[\partial_{m}a^{I}\partial^{m}
\bar{a}^{J}\,+ \, i\bar{\psi}^{J}_{\dot{\alpha}}
\left(\bar{\sigma}^{m}\right)^{\dot{\alpha}\alpha}D_{m}\psi^{I}_{\alpha}\,+\, 
 i\bar{\lambda}^{J}_{\dot{\alpha}}
\left(\bar{\sigma}^{m}\right)^{\dot{\alpha}\alpha}D_{m}\lambda^{I}_{\alpha}\right] 
\nonumber \\ 
{\cal L}_{2} & = & \,{\rm Im}\left[-\frac{1}{2}\F^{(2)}_{IJ}\left(v^{\rm SD}\right)^{I}_{mn}
\left(v^{\rm SD}\right)^{J\,mn} \,\,+\,\, \right. \nonumber \\ 
& & \left. \frac{1}{\sqrt{2}}\F^{(3)}_{IJK} \,
\lambda^{\alpha I}
\left(\sigma^{mn}\right)^{\beta}_{\alpha}\psi^{J}_{\beta}v_{mn}^{K} 
\,\,+\,\, 
\frac{1}{4}\F^{(4)}_{IJKL} \,
\psi^{\alpha I}\psi^{J}_{\alpha}\lambda^{\beta
  K}\lambda_{\beta}^{L}\right]  
\nonumber \\
{\cal L}_{3}  & = &
-\im \tau_{IJ}\left[F^{I}\bar{F}^{J}\,+\,\frac{1}{2}D^{I}D^{J}\right]
\nonumber \\ &&
-\frac{1}{2}{\rm Im}\left[\F^{(3)}_{IJK}\left(\bar{F}^{I}
\left(\psi^{\alpha J}\psi_{\alpha}^{K}+ \lambda^{\alpha J}
\lambda_{\alpha}^{K}\right)\,-\, i\sqrt{2}D^{I}\psi^{\alpha
J}\lambda_{\alpha}^{K}\right) \right]. 
\nonumber 
\end{eqnarray}
Here we have introduced complex and real auxiliary fields $F^{I}$ and
$D^{I}$ respectively which can be eliminated using their equations of
motion. The supersymmetry transformations for this action can be found
in \cite{Dorey:Khoze:Mattis:1996}. 
\paragraph{}
We now consider the dimensional reduction of this action to $0+1$
dimensions by setting 
\begin{eqnarray}
\partial_{m} & = & \frac{\partial}{\partial t} \qquad{} \qquad{} m=0
\nonumber \\ 
& = & 0 \qquad{} \qquad{} m=1,2,3. 
\nonumber 
\end{eqnarray}
The surviving
fields are the scalars $a^{I}$, $\bar{a}^{I}$, the fermions
$\psi^{I}_{\alpha}$, $\lambda^{I}_{\alpha}$,
$\bar{\psi}^{I}_{\dot{\alpha}}$,  $\bar{\lambda}^{I}_{\dot{\alpha}}$,
the auxiliary fields $F^{I}$, $\bar{F}^{I}$, $D^{I}$ and the electric
field strength $E^{I}_{l}=v_{0l}^{I}$ for $l=1,2,3$. 
\paragraph{}
The reduction breaks the
four-dimensional Lorentz group down to three-dimensional spatial
rotations denoted $SU(2)_{N}$ and the reduced theory has a manifest
bosonic symmetry group 
\begin{eqnarray}
& SU(2)_{N}\times SU(2)_{R}\times U(1)_{R}. & 
\nonumber 
\end{eqnarray}
Less obviously the fields of the reduced theory can be combined into
multiplets of an $SU(4)_{R}$ which contains $SU(2)_{N}\times
SU(2)_{R}$ as a subgroup. The  $SU(2)_{N}\times
SU(2)_{R}$ quantum numbers of the various surviving
fields and their lift to $SU(4)_{R}$ are given in the table below: 
\begin{center}
\begin{tabular}{ l| c c| c } 
   & $SU(2)_{N}$ & $SU(2)_{R}$ & $SU(4)_{R}$  \\ \hline
$a$, $\bar{a}$  & ${\bf 1}$ & ${\bf 1}$ & ${\bf 1}$ \\ 
$\psi$, $\lambda$  & ${\bf 2}$ & ${\bf 2}$ & ${\bf 4}$ \\ 
$\bar{\psi}$, $\bar{\lambda}$  & ${\bf 2}$ & ${\bf 2}$ & $ \bar{\bf
    4}$  \\ $E$ & ${\bf 3}$ & ${\bf 1}$ & \multirow{2}{*}{${\bf 6}$} \\
$F,\bar{F},D$ & ${\bf 1}$ & ${\bf 3}$ &
\end{tabular}
\end{center}
As indicated in the table the fermion components can easily be
assembled into a ${\bf 4}$ and $\bar{\bf 4}$ of $SU(4)_{R}$. 
The electric field strength $E$ and
auxiliary fields are combined to form a bosonic field $\vec{\chi}$
transforming in the ${\bf 6}$ of $SU(4)_{R}\simeq
SO(6)_{R}$. Explicitly we form an $SO(6)$ vector 
\begin{eqnarray}
\vec{\chi} & = & \left(\begin{array}{c} E_{1} \\ 
E_{2} \\ E_{3} \\ D \\ \sqrt{2}{\rm Re}[F] \\ \sqrt{2}{\rm
  Im}[F]\\ \end{array}\right). 
\nonumber 
\end{eqnarray}
The vector ${\bf 6}$ of $SO(6)$ corresponds to a second rank
pseudo-real antisymmetric tensor representation of $SU(4)$. 
The map between these
representations involves a vector $\vec{\Sigma}_{AB}$ of $4\times 4$
anti-symmetric matrices. One possible choice is 
\begin{eqnarray}
\vec{\Sigma} & = & (\eta^{1}, \eta^{2}, \eta^{3}, i\bar{\eta}^{1}, 
i\bar{\eta}^{2}, i\bar{\eta}^{3}), 
\nonumber
\end{eqnarray}
where $\eta^{a}_{AB}$ and $\bar{\eta}^{a}_{AB}$ are the 't Hooft
symbols corresponding to self-dual and anti-self-dual generators of
$SO(4)$ respectively. Thus we define a complex anti-symmetric tensor
field $\chi_{AB} =  \vec{\chi}\cdot \vec{\Sigma}_{AB}$ which obeys 
the pseudo-reality condition 
\begin{eqnarray}
\bar{\chi}^{AB} & = & \frac{1}{2}\epsilon^{ABCD}\chi_{CD}. 
\nonumber
\end{eqnarray}
In summary, we now introduce new fields
\begin{eqnarray}
& a^{I}, \qquad{} \psi^{IA}, \qquad{} \chi^I_{AB} & \nonumber \\ 
& \bar{a}^{I}, \qquad{} \bar{\psi}^{I\bar{A}}, \qquad{}
\bar{\chi}^I_{\bar{A}\bar{B}} &
\nonumber 
\end{eqnarray} 
in the $({\bf 1}\oplus{\bf 4}\oplus {\bf 6})\oplus
({\bf 1}\oplus \bar{\bf 4}\oplus {\bf 6})$ of $SU(4)_{R}$.
\paragraph{}
The Lagrangian of the reduced theory can be written in a
manifestly $SU(4)_{R}$-invariant form  
\begin{eqnarray}   
{\cal L} & = & \im \tau_{IJ} \left[\dot{a}^{I}\dot{\bar{a}}^{J}\,+\,
i\bar{\psi}^{J}_AD_{t}\psi^{IA}\,+\,
\chi^I_{AB}\bar{\chi}^{JAB}\right] 
\nonumber \\ & +& \frac{1}{4\pi}{\rm Im}\left[\frac{1}{\sqrt{2}} 
\F^{(3)}_{IJK} \chi^I_{AB}\psi^{JA}\psi^{KB}
\,+\, \frac{1}{48i}\F^{(4)}_{IJKL}
\epsilon_{ABCD}\psi^{IA}\psi^{JB}\psi^{KC}\psi^{LD} \right].
\nonumber 
\end{eqnarray} 
In the above the time derivatives of $\chi_{AB}$ do not
appear and it can be treated as an auxiliary field. This may seem a
little odd as three of the six independent components of $\chi_{AB}$ 
started life as electric field strengths in four dimensions and are
naturally thought of as time derivatives of a vector
potential. However, this is consistent after our dimensional reduction
where corresponding spatial derivatives of the vector potential are
set to zero. Finally 
to make contact with theory of the previous section,
we integrate out the auxiliary fields to get the following Lagrangian 
\begin{eqnarray}   
{\cal L} & = & \im\tau_{IJ}
\dot{a}^{I}\dot{\bar{a}}^{J}\,+\,
i\im\tau_{IJ}
\bar{\psi}^J_AD_{t}\psi^{IA} \nonumber \\ & & \quad{} \quad{}  
\quad{} \quad{} \,-\,
\frac{1}{12}
{\rm Re}\left(\epsilon_{ABCD}{G}_{IJKL}
\psi^{IA}\psi^{JB}\psi^{KC}\psi^{LD}\right)\,-\,
\frac{1}{2}\,R_{I\bar{J}K\bar{L}}\psi^{IA}\bar{\psi}^{\bar{J}}_{A}\psi^{KB}
\bar{\psi}^{\bar{L}}_{B},
\nonumber 
\end{eqnarray} 
where $R_{I\bar{J}K\bar{L}}$ and $G_{IJKL}$ are as in
(\ref{eqn:baseRiemann}) and (\ref{eqn:chiraltensor})
. Performing a Legendre transform on the
above Lagrangian, we arrive at the $SU(4)\simeq SO(6)$-invariant
Hamiltonian (\ref{eqn:truncatedHamiltonian}) of the previous section.

\section{Discussion}
In this paper we have shown that a class of quantum mechanical $\sigma$-models
with scale-invariant special K\"{a}hler target space have $SU(1,1|4)$ 
superconformal invariance. 
The Coulomb branches of four-dimensional superconformal
field theories with ${\cal N}=2$ supersymmetry provide a large class of
examples. We also study the related quantum mechanical $\sigma$-model on 
the Coulomb branch of the
same ${\cal N}=2$ theories compactified down to three dimensions on a
circle of radius $R$. 
The DLCQ description of the ${\cal N}=4$ theory arises as a special
case corresponding to the four-dimensional $\hat{A}_{N-1}$ quiver
described in Section 3. 
Each of these models is
determined by the same data as the corresponding Seiberg-Witten 
solution, namely a complex curve $\Sigma$ with holomorphic differential
$\lambda$. These in turn specify the holomorphic 
prepotential ${\cal  F}$, which determines the Hamiltonian and the 
other superconformal symmetry generators explicitly.  
The simplest possible model one could consider corresponds
to the classical prepotential for a simple gauge group of rank $r$ 
\begin{eqnarray}
{\cal F} & = &\sum_{I=1}^{r}\, \frac{1}{2}\tau a_{I}^{2}.
\nonumber 
\end{eqnarray}
In this case at least, the target space has only orbifold
singularities coming from the fixed points of the Weyl group and the
superconformal symmetry generators are globally defined.   
\paragraph{}
The situation becomes more interesting as soon as quantum corrections
to the Coulomb branch metric are included. These famously lead to
singular submanifolds on which charged BPS states of the
four-dimensional ${\cal N}=2$ theory become masslessy\footnote{A
  related phenomenon in the quantum mechanics model 
is that the corresponding cycles of the
Jacobian torus decompactify at these points and states with momentum
in the fibre directions fail to decouple as $R\rightarrow \infty$.}. 
For a model of
rank one, in suitable local coordinates, the prepotential has 
the characteristic form 
\begin{eqnarray}
{\cal F} & \sim & a^{2}\log a. 
\nonumber 
\end{eqnarray}  
This causes at least two problems. First, the
corresponding target space metric is singular at $a=0$. This means
that quantum mechanics on the manifold is not obviously well-defined,
at least for states with wavefunctions supported near the
singularity.
It is therefore necessary to regulate the model by
resolving the singularity. In the present case a natural regulator is
obtained by working with the full hyper-K\"{a}her model of Section 
\ref{section:sigmamodel} at a finite
value of the compactification radius $R$. Although the semi-flat
metric (\ref{eqn:Gsf}) we have studied in this paper is also
singular, it is known \cite{Gaiotto:Moore:Neitzke:2008} 
that the logarithmic singularities discussed
above are resolved by instanton 
corrections\footnote{In the superconformal models of interest here, singularities of higher
codimension persist even at finite $R$. In the Higgs branch description, 
these are essentially the familiar singularities associated with small
instantons on $\mathbb{R}^{4}$. The standard approach to their resolution
involves the introduction of spacetime non-commutativity \cite{Nekrasov:Schwarz:1998}.} for any
finite value of $R$. Of course the same instanton effects will also
break the superconformal invariance of the model explicitly.  
\paragraph{}
The second, related, problem introduced by the logarithmic singularity 
discussed above is that the superconformal generators themselves are
no longer single-valued on the Coulomb branch. The special conformal
generator $K$, in particular, depends on the first derivative of the
K\"{a}hler potential which has non-trivial monodromy around the
singular point. This means that we can only define the action of the
superconformal group locally 
in some patch which does not contain a singular point. Alternatively
one could consider globally defined generators acting 
on an infinitely-branched cover of the target space. These issues and
their implications for the DLCQ of the ${\cal N}=4$ theory will
be revisited in a forthcoming paper \cite{ND:AS:future}.   

\[ \scriptstyle{****************************}\]
The research leading to these results has received funding from the 
European Research Council under the European Community's Seventh 
Framework Programme (FP7/2007-2013) / ERC grant agreement no. [247252]. 
AS is supported by the STFC.        

\begin{appendices}
\section{Connection and curvature of
  $T^*\M$} \label{appendix:geometry}
In this appendix we give the connection and curvature forms for the
 Levi-Civita connection corresponding to the metric (\ref{eqn:Gcotangent}).
 We work in the bases (\ref{eqn:horizontal}) and (\ref{eqn:coframelift})
 determined by horizontal lifting and use the notational conventions
 outlined at the beginning of section \ref{section:sigmamodel}. The
 connection components are defined by
\[ \nabla_{e_A}e_B = \theta^C_{~BA}e_C\]
where $e_A$ is a generic frame vector, and 
the nonzero components are
\begin{align} \label{eqn:connection} \begin{aligned} \theta^K_{~JI} &=
 -\frac{i}{2}\inv^{KL}\F^{(3)}_{IJL} \\ \theta^{KJ'I'} &= \frac{i}{2}\F^{(3)}_{LMN}
 \inv^{KL} \inv^{J'M}\inv^{I'N} \\ \theta_{K'J}^{~~\bar{I}'} &=
 \frac{i}{2}\F^{(3)}_{JK'L}\inv^{L\bar{I}'} \\ \theta_{K' ~ I}^{~~J'}
 &= \frac{i}{2}\F^{(3)}_{IK'L}\inv^{J'L}. \end{aligned} \end{align}
The fact that the components with mixed (anti)holomorphic
 indices do not vanish is perfectly consistent since the frame vectors $D_I$
 are not holomorphic.
\paragraph{}
The curvature forms are defined by
\[ \Omega^A_{~B} = d\theta^A_{~B} + \theta^A_{~C}\wedge \theta^C_{~B}.\]
To construct the action (\ref{eqn:genericaction}) we need the curvature
 tensor with all lowered indices. The nonzero components of this are
\begin{align} \label{eqn:curvature} \begin{aligned} \Omega_{\bar{I}JK\bar{L}}
 &= R_{\bar{I}JK\bar{L}} \\ \Omega_{\bar{I}J}^{~~~K'\bar{L}'} &= 
R_{\bar{I}J}^{~~~K'\bar{L}'} \\ \Omega^{\bar{I}'J'K'\bar{L}'} &= 
R^{\bar{I}'J'K'\bar{L}'} \\ \Omega_{\bar{I}~K}^{~J'~\bar{L}'} &= 
R^{\bar{L}'J'}_{~~~K\bar{I}} \end{aligned} && \begin{aligned} 
\Omega^{\bar{I}'~~~\bar{L}'}_{~JK} &= -\inv^{\bar{I}'N}\inv^{\bar{L}'M}G_{JKMN} \\ \Omega^{\bar{I}'~~~L'}_{~J\bar{K}} &= R^{\bar{I}'~L'}_{~\bar{K}~J}
 \\ \Omega^{\bar{I}'J'}_{~~~K\bar{L}} &= R^{\bar{I}'J'}_{~~~K\bar{L}}
 \\ \Omega_{\bar{I}~K}^{~J'~L'} &=
 \inv^{J'\bar{M}}\inv^{L'\bar{N}}\bar{G}_{\bar{I}\bar{K}\bar{M}\bar{N}}
  \end{aligned} \end{align} 
along with conjugates and terms obtained via the trivial
 symmetry $\Omega_{ABCD} = -\Omega_{ABDC}$. The tensors
$R_{I\bar{J}K\bar{L}}$ and $G_{IJKL}$ are as in
(\ref{eqn:baseRiemann}) and (\ref{eqn:chiraltensor}).

\section{Summary of $SU(1,1|4)$} \label{appendix:su114}
$SU(1,1|4)$ is a simple superalgebra with bosonic part
 $SO(2,1)\times U(4)$ and may be represented in terms of
 $(2|4)\times (2|4)$ supermatrices
\[ \left(\begin{array}{c|c} SL(2;\mathbb{R}) & \mbox{fermions}
 \\ \hline \mbox{fermions} & SU(4)\end{array} \right)\]
with a diagonal $U(1)$ factor satisfying $\mbox{Str}=0$.
\paragraph{}
The charges generating $SU(1,1|4)$ are
\begin{align} \nonumber \begin{aligned} H &=
 \inv^{I\bar{J}}\Pi_I\bar{\Pi}_{\bar{J}} + \frac{1}{2}R_{I\bar{J}K\bar{L}}\psi^{IA}\bar{\psi}^{\bar{J}}_A\psi^{KB}\bar{\psi}^{\bar{L}}_B
 &&\\&\quad+ \frac{1}{12}\re \left(\epsilon_{ABCD}G_{IJKL}\psi^{IA}\psi^{JB}\psi^{KC}\psi^{LD}\right)
 &&\\ D &= a^I\Pi_I + \bar{a}^{\bar{I}}\bar{\Pi}_{\bar{I}}
 \\ K &= \im \left(\frac{\partial \F}{\partial a^I}\bar{a}^I\right)
 &&\\ \mathcal{R} &= i\left(a^I\Pi_I - \bar{a}^{\bar{I}}\bar{\Pi}_{\bar{I}}\right)
 + \frac{1}{2}\im \tau_{I\bar{J}}\psi^{IA}\bar{\psi}^{\bar{J}}_A
 &&\\ R^{A\bar{B}} &=
 i\im \tau_{I\bar{J}}\left(\psi^{IA}\bar{\psi}^{\bar{J}\bar{B}}
-\frac{1}{4}\delta^{A\bar{B}}\psi^{IC}\bar{\psi}^{\bar{J}}_C\right)
 &&\\ Q^A &= \psi^{IA}\Pi_I + \frac{1}{12}\epsilon^A_{\bar{B}\bar{C}\bar{D}}\bar{\F}^{(3)}_{\bar{I}\bar{J}\bar{K}}
\bar{\psi}^{\bar{I}\bar{B}}\bar{\psi}^{\bar{J}\bar{C}}
\bar{\psi}^{\bar{K}\bar{D}} & \bar{Q}^{\bar{A}} &= 
\left(Q^A\right)^{\dagger} \\ S^A &= \im \tau_{I\bar{J}}\bar{a}^{\bar{J}}\psi^{IA}
 & \bar{S}^{\bar{A}} &= \left(S^A\right)^{\dagger}. \end{aligned} \end{align}
The non-vanishing boson-boson commutators are
\begin{subequations} \nonumber \begin{align} \left[H,K\right]
 &= -iD \qquad \left[D,K\right] = -2iK \qquad \left[D,H\right]
 = 2iH \\ \left[R^{A\bar{B}},R^{C\bar{D}}\right] &= 
i\left(\delta^{C\bar{B}}R^{A\bar{D}} - 
\delta^{A\bar{D}}R^{C\bar{B}}\right). \end{align}
 \end{subequations}
The nonzero fermion charges are
\begin{subequations} \nonumber \begin{align} \left[D,Q^A\right]
 &= iQ^A & \left[D,S^A\right] &= -iS^A \\ \left[\mathcal{R},Q^A\right]
 &= -\frac{1}{2}Q^A & \left[\mathcal{R},S^A\right] &= -
\frac{1}{2}S^A \\ \left[H,S^A\right] &= -iQ^A & \left[K,Q^A\right]
 &= iS^A \\ \left[R^{A\bar{B}},Q^C\right] &= 
i\left(\delta^{C\bar{B}}Q^A - \frac{1}{4}\delta^{A\bar{B}}Q^C\right)
 & \left[R^{A\bar{B}},S^C\right] &= i\left(\delta^{C\bar{B}}S^A
 - \frac{1}{4}\delta^{A\bar{B}}S^C\right)   \end{align}
 \end{subequations}
as well as those following from conjugation. Finally, the
 non-vanishing anticommutators are
\begin{subequations} \nonumber \begin{align} \left\{Q^A,
\bar{Q}^{\bar{B}}\right\} &= \delta^{A\bar{B}}H \qquad
 \left\{S^A,\bar{S}^{\bar{B}}\right\} = \delta^{A\bar{B}}K
 \\ \left\{Q^A,\bar{S}^{\bar{B}}\right\} &=
 \frac{1}{2}\delta^{A\bar{B}}\left(D-i\mathcal{R}\right) -
 R^{A\bar{B}}. \end{align} \end{subequations}

\section{Deformation of $SU(1,1|4)$ for nonzero fibre momentum}
 \label{appendix:deformation}
To obtain $SU(1,1|4)$ invariance it is necessary to truncate to 
the sector with $P^{I'}=0$, both for conformal invariance and
R-symmetry enhancement. One can still define generators at
nonzero fibre momentum, but their algebra no longer closes.
Nevertheless it may be instructive to have explicit expressions
for their generators and commutation relations.
\paragraph{}
The generators which are defined differently are the Hamiltonian
\begin{align} \nonumber \begin{aligned} H &= \inv^{I\bar{J}}\Pi_I\bar{\Pi}_{\bar{J}}
+ \frac{1}{2}R_{I\bar{J}K\bar{L}}\psi^{IA}\bar{\psi}^{\bar{J}}_B\psi^{KB}\bar{\psi}^{\bar{L}}_B
\\ &+ \frac{1}{12}\re \left(\epsilon_{ABCD}G_{IJKL}\psi^{IA}\psi^{JB}\psi^{KC}\psi^{LD}\right)
\\ &+ \im \tau_{I'\bar{J}'}P^{I'}\bar{P}^{\bar{J}'}
+ \frac{1}{2}\re \left( \F^{(3)}_{I'JK}\Omega_{AB}\psi^{JA}\psi^{KB}P^{I'}\right)
\end{aligned} \end{align}
and the supersymmetries
\begin{align} \nonumber \begin{aligned} Q^A &= \psi^{IA}\Pi_I
+ \frac{1}{12}\epsilon^A_{~\bar{B}\bar{C}\bar{D}} \bar{\F}^{(3)}_{\bar{I}\bar{J}\bar{K}}
\bar{\psi}^{\bar{I}\bar{B}}\bar{\psi}^{\bar{J}\bar{C}}\bar{\psi}^{\bar{K}\bar{D}}
\\ &+ \im \tau_{I'\bar{J}}P^{I'}\Omega^A_{~\bar{B}}\bar{\psi}^{\bar{J}\bar{B}}.
\end{aligned} \end{align}
These are just as in (\ref{eqn:invariantHamiltonian}) and (\ref{eqn:SUSYcharges}).
In each case these deformations manifestly break $SO(6) \rightarrow SO(5)$
by use of the symplectic form $\Omega_{AB}$, as well as less
obviously breaking conformal invariance. The latter breaking is
exemplified by the following deformed commutation relations:
\begin{align} \nonumber \begin{aligned} \left[D,H\right] &=
2iH -2i \im \tau_{I'\bar{J}'}P^{I'}\bar{P}^{\bar{J}'} \\
&-\frac{i}{2}\re\left( \F^{(3)}_{I'JK}\Omega_{AB}\psi^{JA}\psi^{KB}P^{I'}\right)
\\ \left[D,Q^A\right] &= iQ^A - i\im \tau_{I'\bar{J}}P^{I'}
\Omega^A_{~\bar{B}}\bar{\psi}^{\bar{J}\bar{B}} \\
\left[H,S^A\right] &= -iQ^A + i \im \tau_{I'\bar{J}}P^{I'}
\Omega^A_{~\bar{B}}\bar{\psi}^{\bar{J}\bar{B}} \\
\left\{Q^A,S^B\right\} &= \im \tau_{I'\bar{J}}P^{I'}\Omega^{AB}\bar{a}^{\bar{J}}.
\end{aligned} \end{align}

\end{appendices}

\bibliography{dlcq-4}
\bibliographystyle{JHEP}

\end{document}